\documentclass[preprint,12pt]{elsarticle}

\usepackage{amssymb}
\usepackage{amsmath}
\usepackage{color}
\usepackage{empheq}

\newcommand{\Fr}{\mathtt{Fr}}  

\renewcommand{\d}{{\rm d}}                
\newcommand{\diag}{{\rm diag}}            

\journal{Advances in Water Resources}

\begin{document}

\begin{frontmatter}




\title{Dam break in rectangular channels with different upstream-downstream widths}

\author[dip]{A. Valiani\corref{Ale}}
\cortext[Ale]{Corresponding author.}
\ead{alessandro.valiani@unife.it}
\author[dip]{V. Caleffi}
\ead{valerio.caleffi@unife.it}
\address[dip]{Universit\`{a} degli Studi di Ferrara,
Dipartimento di Ingegneria, Via G. Saragat, 1 44122 Ferrara, Italy}

\begin{abstract}
The classic Stoker dam-break problem \cite{Stoker} is revisited in cases of different channel widths upstream and downstream of the dam. The channel is supposed to have a rectangular cross section and a horizontal and frictionless bottom. The system of the shallow water equations is enriched, using the width as a space-dependent variable, together with the depth and the unit discharge, which conversely depend on both space and time. Such a formulation allows a quasi-analytical treatment of the system, whose solution is similar to that of the classic Stoker solution when the downstream/upstream depth ratio is sufficiently large, except that a further stationary contact wave exists at the dam position. When the downstream/upstream depth ratio is small, the solution is richer than the Stoker solution because the critical state occurs at the dam position and the solution itself becomes resonant at the same position, where two eigenvalues are null and the strict hyperbolicity of the system is lost. The limits that identify the flow regime for channel contraction and channel expansion are discussed after showing that the nondimensional parameters governing the problem are the downstream/upstream width ratio and the downstream/upstream initial depth ratio.

After the introduction of the previous analytical framework, a numerical analysis is also performed to evaluate a numerical method that is conceived to suitably capture rarefactions, shock waves and contact waves. A second-order method is adopted, employing a Dumbser-Osher-Toro Riemann solver equipped with a nonlinear path. Such an original nonlinear path is shown to perform better than the classic linear path when contact waves of large amplitude must be captured, being able to obtain specific energy conservation and mass conservation at the singularity.

The codes, written in MATLAB (MathWorks Inc.) language, are made available in Mendeley Data repository.
\end{abstract}

\end{frontmatter}


\section{Introduction}
\label{sec:Intro}
The dam-break problem is a classic topic in open channel hydraulics, both for its relevance in dam engineering and for the intrinsic features of interest of this fluid mechanics problem \cite{Stoker, Henderson1966, Liggett, Chaudry}. At the end of the 1990s, a part of the European hydraulic community joined the CADAM project, the European Concerted Action on Dam-Break Modeling project \cite{Cadam}, to take stock of the existing numerical methods and to create a suitable database of laboratory experiments and case studies to improve the technical knowledge and best practices in the field. A relevant conclusion from the CADAM project was that one- and two-dimensional shallow water equations (SWE) can be considered one of the most appropriate tools for dam-break flow modeling on real valleys \cite{Valiani2002, Caleffi2003}. Further reviews can be found in \cite{Singh2011, Wang2011} and the references therein.

Research into the basic aspects of the flow remains active, also taking into account the new insights on the role of the source terms in the SWE, particularly those concerning discontinuous solutions over geometric singularities. The first kind of considered singularity is the discontinuous bottom profile, as studied in \cite{alcrudo01}, which described the standing wave on a bottom step in a SWE Riemann problem where initial depth, velocity and bottom elevation are discontinuous at the same position, showing the existing variety of solutions and the unit discharge and total head conservation at the bottom step. The associated energy dissipation at the step can be taken into account by adding a head loss proportional to the kinetic energy of the flow via an empirical coefficient deduced by selected experiments.

The latter is an important point because, as extensively shown and discussed in \cite{Valiani2017}, the head losses at the singularity are not intrinsically included in the shallow water scheme but must be separately modeled with an additional term that is added to the distributed head loss due to bottom friction.

A further analytical study on bed discontinuities is \cite{LeFloch2011}, where some ideas proposed in \cite{LeFloch2007} are developed. In \cite{LeFloch2011}, the Riemann problem over an uneven bed elevation is studied, proving the existence and possible multiplicity of the solution. The occurrence of such multiplicity in the resonant regime is deeply investigated. The existence of a stationary wave on the bed discontinuity, which maintains the constancy of the unit discharge and of total head over the same discontinuity, is demonstrated and discussed. While \cite{LeFloch2011} is devoted to understanding the nature of the problem, the recent work \cite{HW2014} investigates all the possible solutions of the Riemann problem on a bottom step, including the drying of the bed in some regions of the flow domain.

Successful applications showing energy conservation at geometric singularities can also be found in \cite{Murillo2013, Murillo2014, Navas2015}. The well-balancing process and an extensive comparison between different numerical approaches of the free surface flow over bottom discontinuities are discussed in \cite{Caleffi2016, Caleffi2017}. The problem is extended, and a solution is proposed, concerning not only well balancing in still water (i.e., the satisfaction of the C-property \cite{BeVa-94}) but also an extension of the balance to steady-state conditions, taking into account the dynamical part of the momentum balance \cite{Caleffi2017}.

Conversely, some aspects concerning the physical meaning of the balance laws are used to conceive a momentum-balancing method for the bottom discontinuities in \cite{Caleffi2009}, while integral aspects are highlighted and discussed in \cite{Valiani2017}. The latter work, examining all the possible steady-state conditions over a backward-facing step and a forward-facing step, shows that the total head conservation and integral momentum balance are completely compatible. This work also shows that an inappropriate integral momentum balance, corresponding to a simplified estimate of the depth on the step, can provide an unphysical gain of energy at the singularity.

In the context of the analytical analysis of the SWE, there are no essential differences between the geometrical singularity consisting of a bed discontinuity and a width change in a rectangular channel. In both cases, it is possible to consider an additional variable to the classic flow variables (flow depth and unit discharge). In the former case, the additional variable is the bottom elevation; in the latter case, it is the channel width. In the case of a bottom discontinuity, the new variable represents the difference between the total head and the specific energy; in the case of width change, the new variable represents the difference between the total discharge and the unit discharge. In both cases, the new introduced variable is a geometrical variable, which experiences a stationary discontinuity. At such a discontinuity, a stationary contact wave occurs, in addition to rarefactions and shocks typical of the classic Stoker dam-break solution. In addition to the classic generalized Riemann invariants, which take place in the channel segments where the bottom elevation or the width are constants, a new couple of generalized Riemann invariants appears, the conservation of the unit discharge and of the total head in the former case, the conservation of the total discharge and of the specific energy in the latter case.

In this work, the classic Stoker solution of the SWE for a flat bottom, in terms of the depth and specific discharge, is considered as stated \cite{Stoker,Liggett}. Then, the channel width is introduced as a further dependent variable, causing a complexity in the general solution, which is by far richer than the original Stoker solution. The two fundamental nondimensional parameters governing the problem are the (downstream/upstream) width ratio and the (downstream/upstream) initial depth ratio. It is shown that in the case of contraction, two possible configurations of the flow after the disappearance of the dam are possible; in the case of expansion, four configurations are possible. The limit curve between the two cases for contraction is analytically found and studied, the same for the three limit curves between the four cases for expansion. In summary, the (width ratio, depth ratio) plane is divided into six regions, each of which is investigated and described.

This work has important analogies and differences with \cite{Cozzolino2018}, which analyzes the behavior of the porous shallow water system, used to study the flow field when a significant part of the domain is covered by buildings, and a simplified technique avoiding a very refined grid between buildings is required. The role of the channel width is similar to the role of the porosity in the urban environment, and the configurations in the case of width changes are strictly similar. From the methodological point of view, the works are quite different because the role of the width is extracted here, generating the third equation (stationarity of the width in time), which allows a simplification of the analytical treatment. In fact, using this approach, the conditions on the standing wave over the discontinuity are automatically recovered from finding the generalized Riemann invariants of the problem. Moreover, all the limit curves are found analytically here, and this may be an added value for classifying various cases. Finally, the numerical method presented herein is completely different from that proposed in \cite{Cozzolino2018}. The \cite{PCL, DLM} general framework for nonconservative systems, within which the DOT method \cite{DOT} is conceived, is considered the ``natural tool'' to solve the augmented shallow water system: the coupling with a proper design of a nonlinear path is the only problem-dependent (bottom elevation steps, abrupt changes in width) ingredient.

A further related work is \cite{Cozzolino2017}, which concerns the Riemann problem in localized constrictions and obstructions in channels, that is, cross-sectional contractions followed by expansions. The context and, consequently, the domain of possible solutions are quite different from the present ones, but several common points can be detected, characterizing the nature of the solution.

For completeness, the extended Stoker problem is also analyzed here from a numerical point of view. The second-order version by \cite{Leibinger} of the so-called DOT (Dumbser-Osher-Toro) method \cite{DOT} is adopted to numerically solve the system governing the problem. This is a generalized path-conservative \cite{PCL} Osher-type Riemann solver for conservative and nonconservative systems. A nonlinear path is proposed to enforce the correct energy and mass conservation at the singularity, inspired by the structure of the generalized Riemann invariant on the contact wave. The method successfully reproduces the Riemann analytical solutions, showing itself to be a relatively simple and highly efficient tool to capture the system behavior at the width change. Finally, conclusions are drawn on the main novelties presented herein.

\section{Governing equations: the augmented shallow water equations system}
\label{sec:GOV}
The original Stoker problem is extended to analyze the behavior of the solution when a sudden contraction or expansion is located at the dam.

The problem is assumed to be governed by the \emph{augmented} shallow water equations system, derived from the original SWE, where the width assumes the role of an additional variable that does not vary with time but can vary and is subjected to abrupt changes with space. This technique has been adopted up to now to capture abrupt changes in the bottom elevation when a constant width is assumed. Such a technique is extensively discussed in \cite{LeFloch2011}.
The \emph{augmented} SWE without friction on a flat bed and varying width in a rectangular cross-section channel can be written as:
\begin{subequations}\label{eq:ESWE}
\begin{align}
&\frac{\partial h}{\partial t} + \frac{\partial q}{\partial x} + \frac{q}{b}\frac{\partial b}{\partial x} = 0
\label{eq:ESWEa}\\
&\frac{\partial q}{\partial t} + \frac{\partial}{\partial x}\left(\frac{1}{2} g h^2 + \frac{q^2}{h} \right) +  \frac{q^2}{bh}\frac{\partial b}{\partial x} = 0
\label{eq:ESWEb}\\
&\frac{\partial b}{\partial t} = 0 
\label{eq:ESWEc}
\end{align}
\end{subequations}
where $h(x,t)$, $q(x,t)$ and $b(x)$ are the depth, the specific discharge and the cross section width, respectively; $g$ is the gravity acceleration; and $x$ and $t$ are the space and the time, respectively.
The compact version of Eq.~\eqref{eq:ESWE} can be written as:
\begin{equation}\label{eq:CSWE}
\frac{\partial W}{\partial t} + A\left(W\right) \frac{\partial W}{\partial x} = 0
\end{equation}
where:
\begin{equation}\label{eq:CESWE}
W=\begin{bmatrix}h\\ q\\b\end{bmatrix}; \quad
A(W)=\begin{bmatrix}0&1&\frac{q}{b}\\g\,h-\frac{q^2}{h^2}&2\frac{q}{h}&\frac{q^2}{b\,h}\\0&0&0\end{bmatrix}
\end{equation}
The column vector $W(x,t)$ contains the evolving variables, and $A\left(W\right)$ is the flux matrix.

\begin{figure}
\begin{center}
\includegraphics[width=1.0\textwidth]{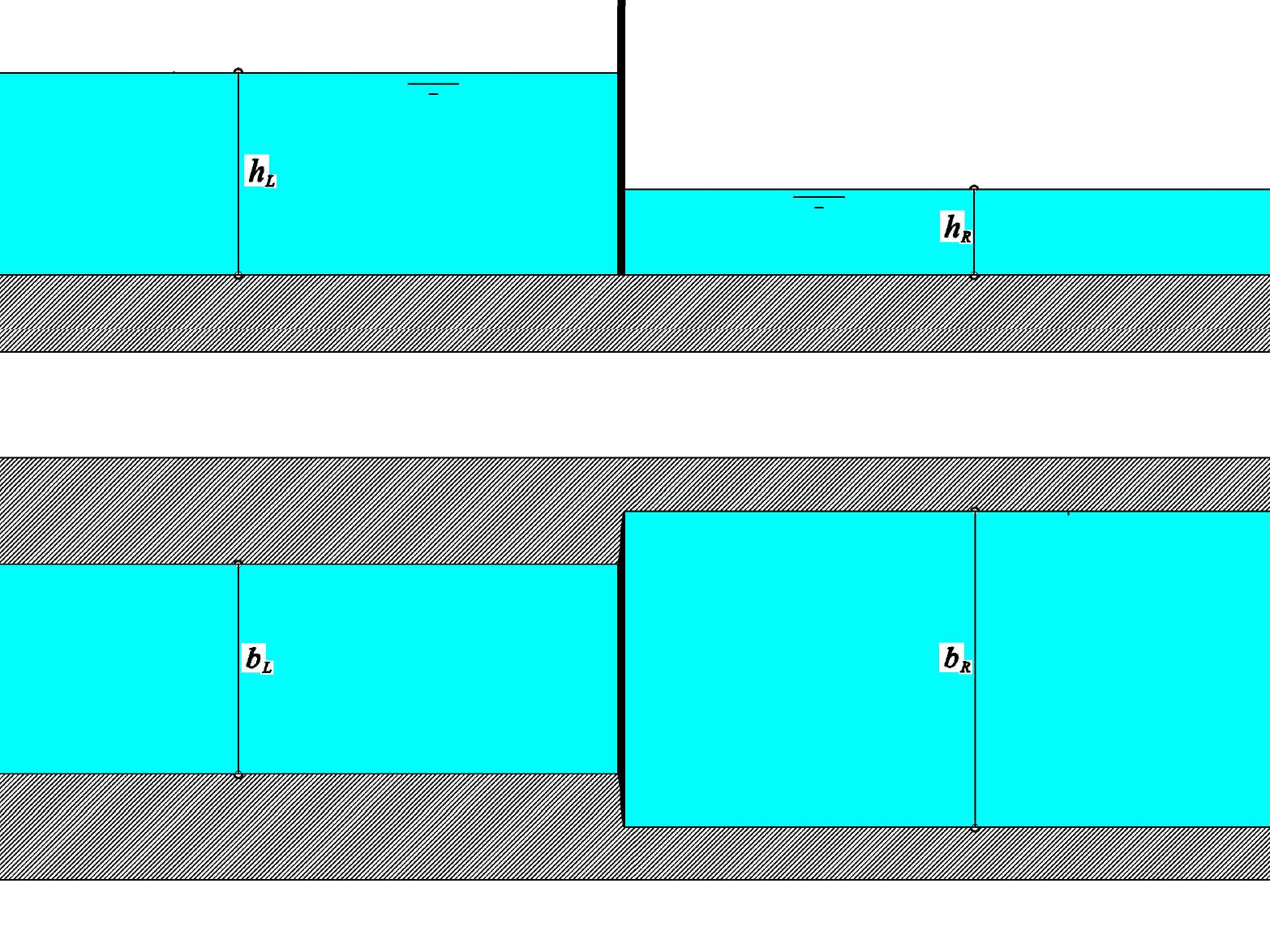}
\end{center}
\caption{Initial conditions for the Stoker dam break in a changing-width channel: in the following, $h_R<h_L$ is always assumed (in the opposite case, symmetric results are obtained); $b_R<b_L$ means contraction, whereas $b_R>b_L$ means expansion.}
\label{fig:widthc}
\end{figure}

The dam divides the channel into two parts where the water elevations are $h_L$ and $h_R$. We assume $h_L>h_R$ (in the opposite case, symmetric results are obtained); therefore, the flow velocities are always positive. The rectangular cross-section channel has two different widths upstream and downstream of the dam position, $b_L$ and $b_R$, respectively. We assume $b_R \gtrless b_L$.
The initial conditions of the extended Stoker problem read:
\begin{equation}\label{eq:BIC}
\begin{bmatrix}h\\ q\\b\end{bmatrix} = \begin{bmatrix}h_L\\ 0\\b_L\end{bmatrix}\quad \text{for} \quad x \le 0 \, ; \quad \text{and} \quad
\begin{bmatrix}h\\ q\\b\end{bmatrix} = \begin{bmatrix}h_R\\ 0\\b_R\end{bmatrix}\quad \text{for}  \quad x> 0
\end{equation}
where the abscissa $x = 0$ indicates the dam position. The extended Stoker problem constituted by the system \eqref{eq:CSWE} with the initial conditions \eqref{eq:BIC} is equivalent to a Riemann problem. The case $b_R=b_L$ coincides with the original Stoker problem \cite{Stoker}.
The sketch of the initial conditions is depicted in Fig.~\ref{fig:widthc}.

The eigenvalues of the flux matrix $A$ are:
\begin{equation}\label{eq:EIG}
\begin{bmatrix}\lambda_1\\  \lambda_2\\ \lambda_3\end{bmatrix} = \begin{bmatrix}u-c\\ 0\\u+c\end{bmatrix}
\end{equation}
where $c = \sqrt{g \, h}$ is the relative celerity of the small-amplitude waves and $u=q/h$ is the depth-averaged velocity.
In the following, the right eigenvectors, the left eigenvectors and the generalized Riemann invariants are considered, in the same order related to the order of the herein-defined eigenvalues.
Note that this is an increasing order if and only if the flow is subcritical ($u<c$); in critical conditions ($u=c$), the strict hyperbolicity of the system is lost, and $\lambda_1=\lambda_2=0$.

The right (normalized) eigenvectors are the columns of the following matrix:
\begin{equation}\label{eq:RIEIG}
R = \begin{bmatrix}1&1&1 \\ u-c&\frac{u^2-c^2}{u}&u+c \\ 0&-\frac{b\left(u^2-c^2\right)}{u^2 \, h}&0\end{bmatrix}
\end{equation}

The left eigenvectors are the rows of the following matrix:
\begin{equation}\label{eq:LEEIG}
L = \begin{bmatrix}+ \frac{u+c}{2 \, c}&- \frac{1}{2 \, c}&+ \frac{u \, h}{2 \, b \left(u-c \right)} \\
0&0&- \frac{u^2 \, h}{b \left(u^2-c^2 \right)} \\
- \frac{u-c}{2 \, c}&+ \frac{1}{2 \, c}&+ \frac{u \, h}{2 \, b \left(u+c \right)}\end{bmatrix}
\end{equation}

The generalized Riemann invariants are:
\begin{subequations}\label{eq:GR13}
\begin{align}
\frac{\d h}{1} &= \frac{\d (u \, h)}{u-c} = \frac{\d b}{0} & &\Rightarrow \left\{ \begin{array}{l} b = const \\ u+2 \, c = const \end{array} \right .  
\label{eq:GR1}\\
\frac{\d h}{1} &= \frac{u}{u^2-c^2}\,\d (u\,h) = -\frac{u^2 \, h}{b \left(u^2-c^2 \right)}\,\d b & &\Rightarrow  \left\{ \begin{array}{l}q\,b = Q = const\\ h + \frac{u^2}{2 \, g} = E = const\end{array} \right .  
\label{eq:GR2}\\
\frac{\d h}{1} &= \frac{\d (u \, h)}{u+c} = \frac{\d b}{0} & &\Rightarrow \left\{ \begin{array}{l} b = const\\ u-2\, c = const\end{array} \right . 
\label{eq:GR3}
\end{align}
\end{subequations}
It is worth noting that $Q$ is the total volumetric discharge and $E$ is the specific energy of the flow. The generalized Riemann invariants \eqref{eq:GR1} and \eqref{eq:GR3} are the ones associated with the classic SWE, while the generalized Riemann invariant \eqref{eq:GR2} is specific to the augmented system and expresses the mass and energy conservation through the standing contact wave at the dam position; we remark that the corresponding eigenvalue $\lambda_2$ is \emph{always} zero.

\section{Analytical solution for channel contraction}
\label{sec:contr}
The Stoker solution for a constant-width channel consists of an unperturbed upstream water level, a rarefaction, a constant state, a moving shock and a still-water downstream state \cite{Stoker, Liggett}. Only two eigenvalues, $u-c$ and $u+c$, exist; the former is associated with the rarefaction, and the latter is associated with the downstream moving shock.

When a width change occurs at the dam position, three eigenvalues exist \eqref{eq:EIG}; the null $\lambda_2$ eigenvalue is associated with a standing contact wave, occurring at $x=0$. A resonant Riemann problem arises when two eigenvalues coincide and become zero at the dam position. Depending on the sign of $u-c$, the order of eigenvalues can commute, in which case the nature of the solution changes.

The first case that is studied here is the contraction, which is $b_R<b_L$. In the following, it is shown that in such cases, two types of solution exist: the former occurs for \emph{large} values of the downstream/upstream depth ratio, $r_h = h_R/h_L$, the latter for \emph{small} values of the same parameter. A \emph{limit} value of the depth ratio divides the two types of solutions. This limit value is not a constant but depends on the downstream/upstream width ratio, $r_b = b_R/b_L$, which influences the nature of the solution, as can be expected on a physical basis.

\subsection{Contraction, large depth ratio}
\label{subsec:CLDR}
First, we consider a \emph{large} downstream/upstream depth ratio, $r_h$, smaller than 1 but sufficiently large to obtain a rarefaction wave that is associated with an everywhere-negative value of the $\lambda_1$ eigenvalue; such a rarefaction wave is located entirely  upstream of the dam position. The intermediate constant state characterizing the original Stoker solution is replaced by two constant states, which are divided by the stationary contact wave located at the dam position; a downstream moving shock divides the downstream constant state from the water at rest corresponding to the right initial state.

\begin{figure}
\begin{center}
\includegraphics[width=0.8\textwidth]{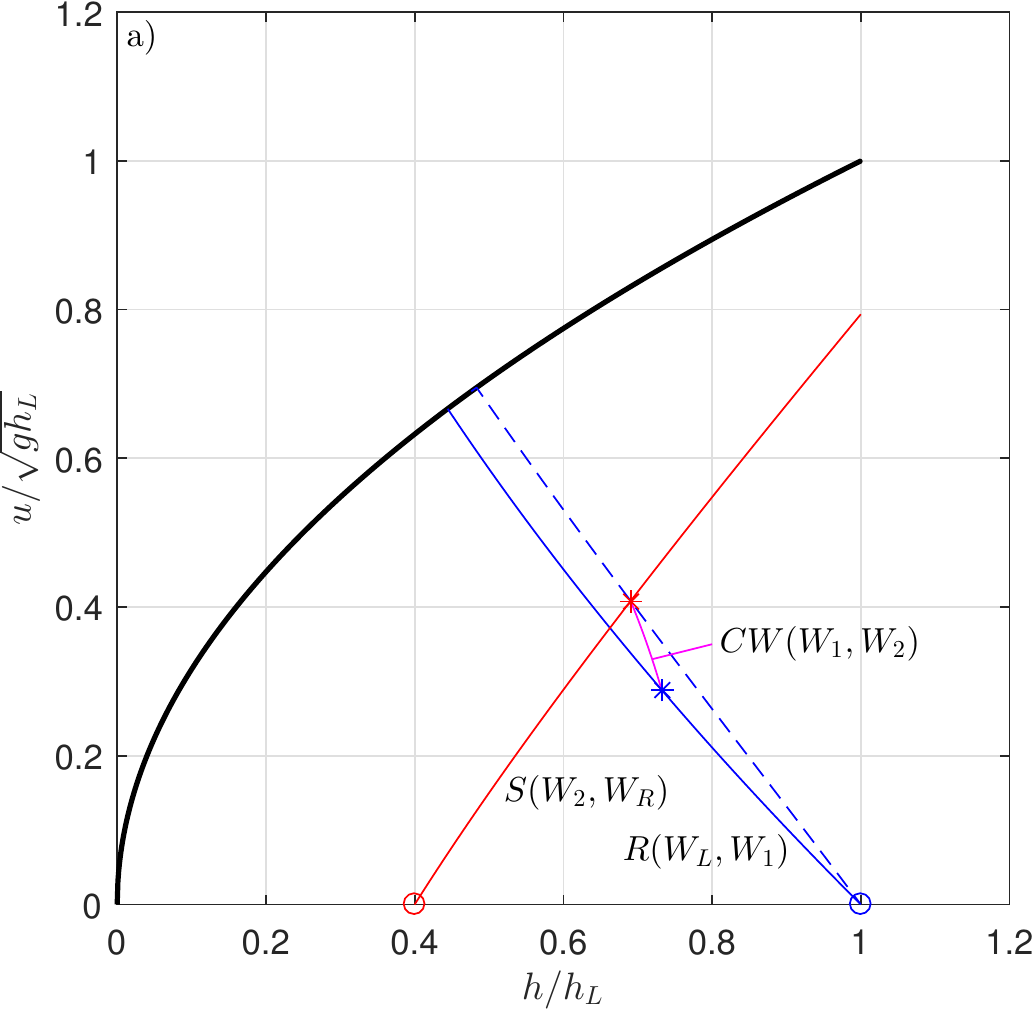}

\rule{0mm}{6mm}

\includegraphics[width=1.0\textwidth]{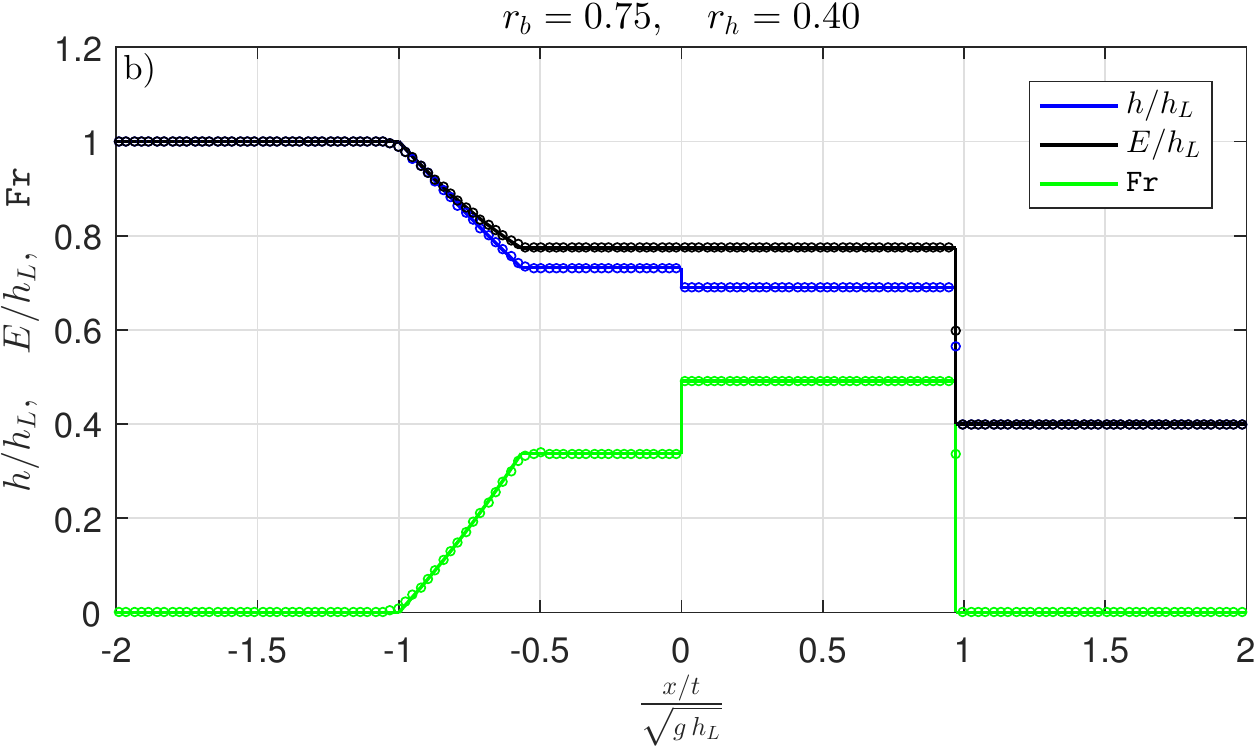}
\end{center}
\caption{Contraction, large $r_h$. a) solution in the phase plane. b) solution in the physical plane. The continuous line is the analytical solution, and the circles represent the numerical solution. }
\label{fig:CLDR}
\end{figure}

Using the classic techniques of  Riemann problem analytical solution, graphically represented in Fig.~\ref{fig:CLDR}a, a rarefaction curve $R$ (continuous blue line) is computed, starting from the left state (blue circle) up to the intersection with the critical resonance curve (thick black line); such a rarefaction curve is referred to as the left channel width $b_L$. A subcritical curve (dashed blue line), imposing the same specific energy and the same total discharge of the rarefaction curve, is drawn; this curve is referred to as the right channel width $b_R$. Starting from the right state (red circle), a shock curve $S$ (continuous red line) is computed, which intersects the dashed blue line at a point (red asterisk). From this point, a constant-energy and constant-discharge contact wave curve $CW$ (continuous magenta line) is computed, considering a linear variation of the channel width from $b_R$ to $b_L$, so that an intersection with the first $R$ curve can be found (blue asterisk). The blue asterisk and the red asterisk identify the two constant states upstream and downstream of the dam, respectively, which appear in the following.

The solution is described by the following system of equations (enumerated moving from upstream to downstream).

First, a constant left state upstream ($x_L$ is the rarefaction head position) is given by:
\begin{equation}\label{eq:CLS}
\left\{ \begin{array}{l} h = h_L \\ u = 0 \end{array} \right . \quad \textrm{for} \quad \frac{x}{t} \leq \frac{x_L}{t}=-\sqrt{g\,h_L}
\end{equation}

The unperturbed left state is followed by a rarefaction associated with the $\lambda_1$ (negative) eigenvalue ($x_1$ is the rarefaction tail position, while $h_1$ and $u_1$ are the corresponding depth and flow velocity, respectively):
\begin{equation}\label{eq:UPRAR}
2\, \sqrt{g\,h_L}=u+2\, \sqrt{g\,h} \quad \textrm{for} \quad \frac{x_L}{t} \leq \frac{x}{t} \leq \frac{x_1}{t}=\left(u_1-\sqrt{g\,h_1} \right)   
\end{equation}
It is worth noting that Eq.~\eqref{eq:UPRAR}, at $x=x_1$, gives:
\begin{equation}\label{eq:UPRAR1}
2\, \sqrt{g\,h_L}=u_1+2\, \sqrt{g\,h_1} \quad \textrm{for} \quad \frac{x}{t} = \frac{x_1}{t}   
\end{equation}

A constant state just upstream of the initial dam position is expressed by the following equations:
\begin{equation}\label{eq:CLSU}
\left\{ \begin{array}{l} h = h_1 \\ u = u_1 \end{array} \right . \quad \textrm{for} \quad \frac{x_1}{t} \leq \frac{x}{t} \leq 0^{-}   
\end{equation}

A constant state just downstream of the initial dam position is expressed by:
\begin{equation}\label{eq:CLSD}
\left\{ \begin{array}{l} h = h_2 \\ u = u_2 \end{array} \right . \quad \textrm{for} \quad 0^{+} \leq \frac{x}{t} \leq \frac{x_2}{t}=c_2  
\end{equation}
where:
\begin{equation}\label{eq:c2}
c_2 = h_2\sqrt{\frac{1}{2}\,g\left(\frac{1}{h_2}+\frac{1}{h_R} \right)}
\end{equation}
is the celerity of the shock (located at $x=x_2$) dividing the constant state downstream of the dam from the constant still-water downstream state.

A constant state, just downstream of the final shock, corresponding to the downstream initial state is given by:
\begin{equation}\label{eq:CRS}
\left\{ \begin{array}{l} h = h_R \\ u = 0 \end{array} \right . \quad \textrm{for} \quad \frac{x}{t} \geq \frac{x_2}{t}   
\end{equation}

At the dam position, $x=0$, mass conservation requires:
\begin{equation}\label{eq:MCD}
u_1\, b_L\, h_1 = u_2\, b_R\, h_2
\end{equation}

Again at $x=0$, energy conservation requires:
\begin{equation}\label{eq:ECD}
h_1+ \frac{u_1^2}{2\,g} = h_2+ \frac{u_2^2}{2\,g}
\end{equation}

At the shock position, $x=x_2$, the Rankine-Hugoniot condition (remembering that $u_R=0$) reads:
\begin{equation}\label{eq:RHx2}
u_2 = \left(h_2-h_R\right) \sqrt{\frac{1}{2}\,g\left(\frac{1}{h_2}+\frac{1}{h_R} \right)}
\end{equation}

Considering the four equations (\ref{eq:UPRAR1}), (\ref{eq:MCD}), (\ref{eq:ECD}), (\ref{eq:RHx2}), a nonlinear system in the four unknowns $(h_1, u_1, h_2, u_2)$ is obtained, which can be solved with a standard Newton-Raphson method.

The typical solution is depicted (continuous line) in the physical plane in Fig.~\ref{fig:CLDR}b, together with the corresponding numerical solution (circles), which will be described in Section \ref{sec:NMCW}. The analytical solution and the numerical solution are drawn together herein and in the following to avoid duplication of each diagram. The behavior of the numerical part will be clarified in Section \ref{sec:NMCW}. To adopt nondimensional variables in Fig.~\ref{fig:CLDR}, the depth and specific energy are scaled with the upstream initial depth; the velocity is scaled with $\sqrt{g \, h_L}$; and the Froude number, $\Fr =u/\sqrt{g \, h}$ is reported instead of the flow velocity.

\subsection{Contraction, small depth ratio}
\label{subsec:CSDR}
We consider here a \emph{small} downstream/upstream depth ratio, $r_h$, which is sufficiently small to obtain a region of the solution where the $\lambda_1$ eigenvalue becomes positive (and, consequently, the order of eigenvalues commutes, and such eigenvalue becomes the second one); this region is located downstream of the dam position. Resonance occurs at the dam position, where two eigenvalues are null. The stationary contact wave located there divides a subcritical state at $0^{-}$ from a critical (sonic) state at $0^{+}$. The downstream (the dam) constant state is supercritical, and a shock moving downstream divides this region from the water at rest corresponding to the right initial state. 

\begin{figure}
\begin{center}
\includegraphics[width=0.8\textwidth]{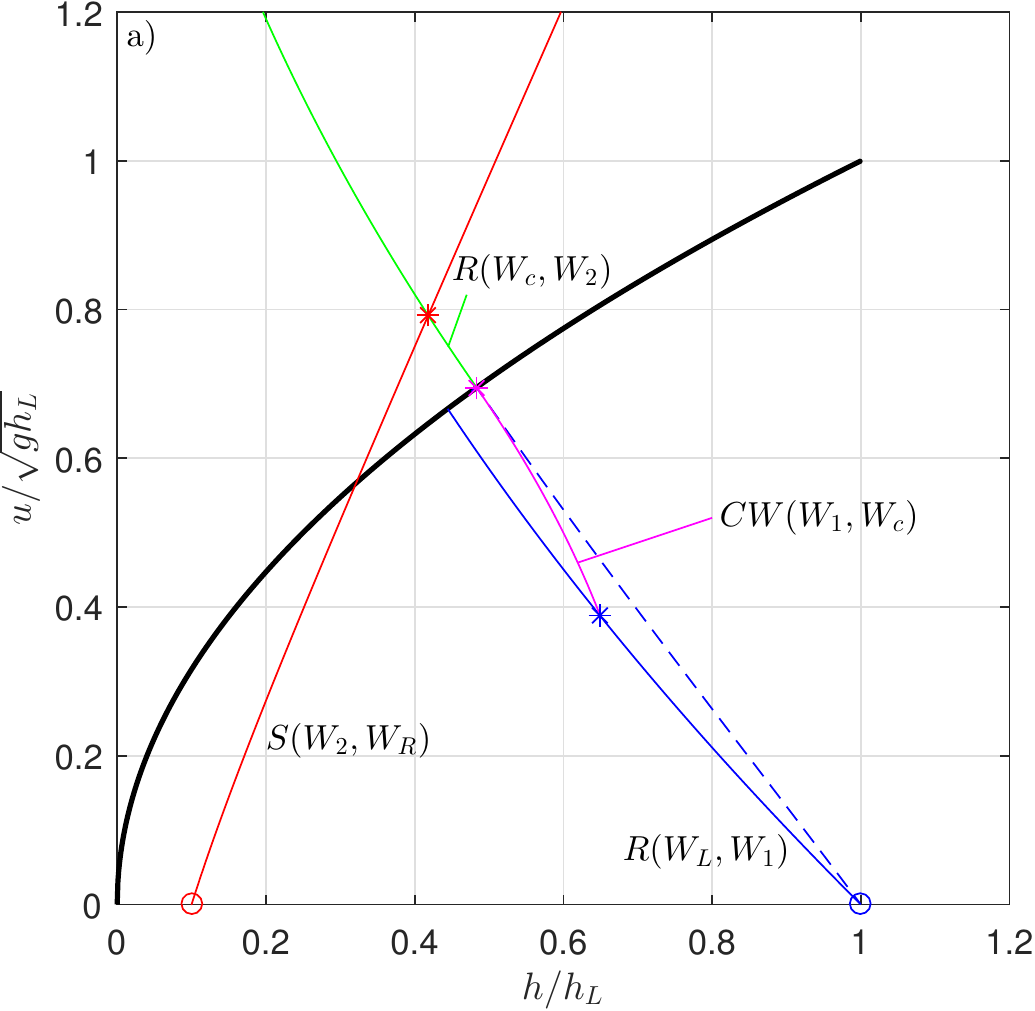}

\rule{0mm}{6mm}

\includegraphics[width=1.0\textwidth]{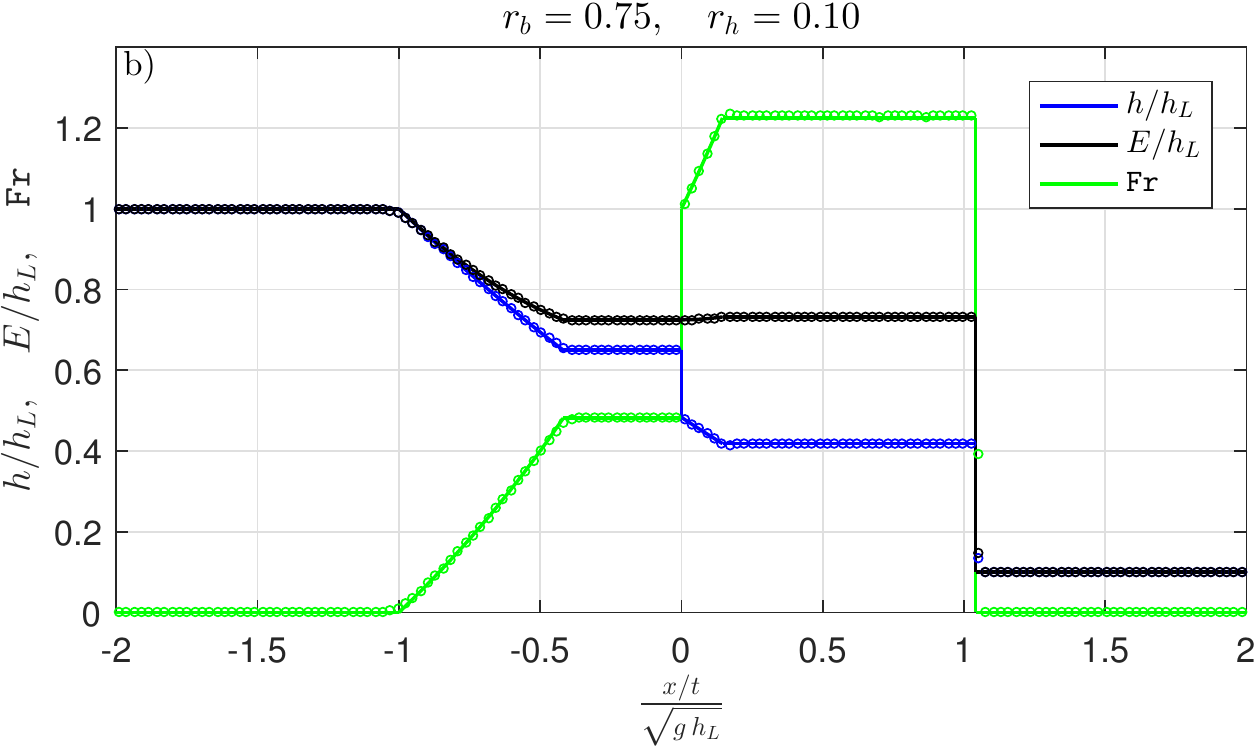}
\end{center}
\caption{Contraction, small $r_h$. a) solution in the phase plane. b) solution in the physical plane. The continuous line is the analytical solution, and the circles represent the numerical solution.}
\label{fig:CSDR}
\end{figure}

Using the classic technique for the Riemann problem analytical solution, graphically represented in Fig.~\ref{fig:CSDR}a), the rarefaction curve $R$ (continuous blue line) is analogous to that of the previous subsection \ref {subsec:CLDR}. The subcritical (dashed blue line) curve, having the same specific energy and the same total discharge of the rarefaction curve, is also analogous to that of the previous subsection \ref {subsec:CLDR}. Starting from the right state, the shock curve (continuous red line) does not intersect the dashed blue line, so that the critical state controls the solution.
The intersection between the dashed blue line and the resonance curve identifies  this critical state (magenta asterisk). From this point, the constant-energy and constant-discharge contact wave curve $CW$ (continuous magenta line) is computed, considering a linear variation of the channel width from $b_R$ to $b_L$, up to the intersection with the first $R$ curve (blue asterisk).
From the critical point a rarefaction curve (continuous green line), in the channel having width $b_L$, is also drawn. This line intersects the shock red curve at a point (red asterisk). 
The blue asterisk and the red asterisk identify the two constant states upstream and downstream of the dam, respectively. The magenta asterisk identifies the critical point at $0^{+}$. The solution is governed by the following system of equations (moving from upstream to downstream).

The critical state of the flow occurs in $x=0^{+}$, that is:
\begin{equation}\label{eq:C0+}
\Fr \left( 0^{+} \right) = 1 \quad \Leftrightarrow \quad u_c=u\left( 0^{+} \right)=\sqrt{g\,h\left( 0^{+} \right)}=\sqrt{g\,h_c}
\end{equation}

The solution is strictly similar to the previous case for the upstream still-water zone, see Eq.~\eqref{eq:CLS}, and for the following rarefaction, see Eq.~\eqref{eq:UPRAR}; the sonic point occurs at $x=0^{+}$; a further rarefaction, associated with the eigenvalue $\lambda_1 = u-\sqrt{g\,h}>0$, occurs downstream of the dam position ($x_{20}$ is the tail of this latter rarefaction):
\begin{equation}\label{eq:DRAR}
3 \sqrt{g\,h_c} = u+2\sqrt{g\,h} \quad \textrm{for} \quad 0 \leq \frac{x}{t} \leq \frac{x_{20}}{t} = \left( u_2 - \sqrt{g\,h_2} \right) 
\end{equation}
and, in particular:
\begin{equation}\label{eq:DRAR20}
3 \sqrt{g\,h_c} = u_2+2\sqrt{g\,h_2} \quad \textrm{for} \quad  \frac{x}{t} = \frac{x_{20}}{t}   
\end{equation}

At the dam position, $x=0$, mass and energy conservation require:
\begin{subequations}
\label{eq:MECDa+b}
\begin{align}
&u_1\, b_L\, h_1 = \sqrt{g\,h_c}\, b_R\, h_c
\label{eq:MECDa}\\
&h_1+ \frac{u_1^2}{2\,g} = \frac{3}{2}h_c
\label{eq:MECDb}
\end{align}
\end{subequations}

A constant state, just downstream of the previously defined rarefaction, exists:
\begin{equation}\label{eq:CLSD1}
\left\{ \begin{array}{l} h = h_2 \\ u = u_2 \end{array} \right . \quad \textrm{for} \quad \frac{x_{20}}{t} \leq \frac{x}{t} \leq \frac{x_2}{t}=c_2   
\end{equation}
where $c_2$ is given by Eq.~\eqref{eq:c2} and relation \eqref{eq:RHx2} still holds. The final constant state zone is given by Eq.~\eqref{eq:CRS}.

In such a case, Eq.~\eqref{eq:UPRAR1}, Eq.~\eqref{eq:DRAR20}, Eqs.~\eqref{eq:MECDa+b}, Eq.~\eqref{eq:RHx2} give a system of five equations in the five unknowns $u_1, h_1, h_c, u_2, h_2$. From Eq.~\eqref{eq:UPRAR1}, $u_1$ is expressed as a function of $h_L$ and $h_1$; substituting in the first and second parts of Eqs.~\eqref{eq:MECDa+b}, we obtain:
\begin{subequations}
\label{eq:interm}
\begin{align}
&h_c \, \sqrt{g h_c}=\frac{b_L}{b_R} \left( 2\,\sqrt{g\,h_L} - 2\,\sqrt{g\,h_1} \right) h_1
\label{eq:interma}\\
&\frac{3}{2}\, h_c = h_1 + \left(2\, h_L + 2\, h_1 - 4 \sqrt{h_1\,h_L} \right)
\label{eq:intermb}
\end{align}
\end{subequations}

The critical depth $h_c$ can be obtained from both Eqs.~\eqref{eq:interm} and then eliminated, obtaining one nonlinear equation, where both sides represent the critical depth $h_c$ and where $h_1$ is the only unknown:
\begin{equation}\label{eq:h1only}
2\, h_1 + \frac{4}{3}\,h_L - \frac{8}{3}\, \sqrt{h_L\, h_1} = \left[ \left( 2\sqrt{h_L}-2\sqrt{h_1} \right) h_1\, \frac{b_L}{b_R} \right]^{2/3}   
\end{equation}
Equation \eqref{eq:h1only} can be solved using the Newton method. Then, Eq.~\eqref{eq:UPRAR1} gives $u_1$, and the following \eqref{eq:hc_h1} gives $h_c$:
\begin{equation}\label{eq:hc_h1}
h_c=2\, h_1 + \frac{4}{3}\,h_L - \frac{8}{3}\, \sqrt{h_L\, h_1}
\end{equation}

It is worth spending more time on Eq.~\eqref{eq:hc_h1}, which expresses, using the left still-water depth $h_L$ as the proper vertical length scale, the link between ${r_c}={h_c}/{h_L}$, the nondimensional critical depth, and ${r_1}={h_1}/{h_L}$, the nondimensional depth just upstream of the dam position at $x=0^{-}$. Eq.~\eqref{eq:hc_h1} can be transformed into a second-degree equation (as evidenced by the position $z=\sqrt{r_1}$) as follows:
\begin{equation}\label{eq:h1ad}
2\, r_1 - \frac{8}{3}\, \sqrt{r_1} + \left(\frac{4}{3} - r_c \right) = 0
\end{equation}

The solutions of Eq.~\eqref{eq:h1ad} are as follows:
\begin{subequations}
\label{eq:phi12}
\begin{align}
r_1^{(1)} &= \left( \frac{2}{3} - \frac{\sqrt{2}}{2}\,\sqrt{r_c - \frac{4}{9}\,}\right)^2
\label{eq:phi1}\\
r_1^{(2)} &= \left( \frac{2}{3} + \frac{\sqrt{2}}{2}\,\sqrt{r_c -  \frac{4}{9}\,}\right)^2
\label{eq:phi2}
\end{align}
\end{subequations}

which are real if and only if:
\begin{equation}\label{eq:R vincM}
r_c \ge \frac{4}{9} \quad \Rightarrow \quad h_c \ge \frac{4}{9}\,h_L.
\end{equation}
To define the selection criteria between the solutions $r_1^{(1)}$ and $r_1^{(2)}$, each side of \eqref{eq:MECDb} is divided by $h_1$, obtaining:
\begin{equation}\label{eq:energy_ad}
1 +  \frac{1}{2} \, \Fr_{1}^{2} = \frac{3}{2}\, \frac{h_c}{h_1}.
\end{equation}
To properly join the upstream boundary condition, the flow at $x=0^{-}$ is subcritical, i.e., $\Fr_{1}\leq1$, and therefore the right-hand side of Eq.~\eqref{eq:energy_ad} is less than or equal to $3/2$. After straightforward manipulation, this requirement can be expressed as $h_1\ge h_c$ and finally as $r_1\ge r_c$. Given Eq.~\eqref{eq:phi2}, $r_1^{(2)}\ge r_c$ and $r_1^{(2)}$ is a physically based solution, while for \eqref{eq:phi1}, we have $r_1^{(1)} < r_c$, and therefore $r_1^{(1)}$ must be discarded. In the following, we indicate with $r_1$ the selected solution.

The latter solution Eq.~\eqref{eq:phi2} gives the value
$r_1=h_1(0^{-})/{h_L}$,
corresponding to the critical state at $x=0^{+}$.

A further constraint to guarantee the physical meaning of the computations is that $r_1 < 1$, to maintain always $h_1<h_L$:
\begin{equation}\label{eq:R vincm}
r_c < \frac{2}{3} \quad \Rightarrow \quad h_c < \frac{2}{3}\,h_L
\end{equation}

It is simple to verify that in the range ${4}/{9} \le r_c < {2}/{3}$, the corresponding range for $r_1$ is ${4}/{9} \le r_1 < 1$, which are the only  physically admissible results in the case of width contraction.

Eq.~\eqref{eq:h1only} and Eq.~\eqref{eq:hc_h1} can also be used to obtain a relationship between $r_b$, $r_c$ and $r_1$:
\begin{equation}\label{eq:rb_ext}
r_b \left( \frac{h_c}{h_L}\right)^{3/2} = \left( 2 - 2 \sqrt{\frac{h_1}{h_L}}\right)\left(\frac{h_1}{h_L}\right)
\end{equation}

Therefore, for this kind of solution, a further relationship between $r_b$, $r_c$ and $r_1$ is as follows:
\begin{equation}\label{eq:rb}
r_{b} \, r_c^{3/2} = 2\, r_{1} \left(1 - \sqrt{r_1}\right)
\end{equation}

Finally, using the rarefaction relation Eq.~\eqref{eq:DRAR20} and Eq.~\eqref{eq:RHx2} for $u_2$, the depth $h_2$ can be found from the Rankine-Hugoniot condition:
\begin{equation}\label{eq:h2_oneq}
3 \, \sqrt{g\, h_c} - 2 \, \sqrt{g\, h_2}=\left(h_2-h_R\right) \sqrt{\frac{1}{2}\,g\left(\frac{1}{h_2}+\frac{1}{h_R} \right)}
\end{equation}
This single equation in only one unknown, $h_2$, can be solved using the Newton method. Finally, the shock celerity $c_2$ is found from Eq.~\eqref{eq:c2}.

The typical solution is depicted (continuous line) in the physical plane in Fig.~\ref{fig:CSDR}b, together with the corresponding numerical solution (circles).

\subsection{Contraction, limit depth ratio}
\label{subsec:CLLDR}
The target of this subsection is to find analytically the limit dividing the \emph{small} downstream/upstream depth ratio from the \emph{large} one to identify the ranges of existence of both. Inspecting the structures of the flow field described in the previous subsections \ref {subsec:CLDR} and \ref {subsec:CSDR}, the limit is recognized to satisfy the equalities $h_2=h_c$, $u_2=u_c$.
The resonance at the dam position still occurs, and this sonic state coincides with the constant state just upstream of the final shock.

\begin{figure}
\begin{center}
\includegraphics[width=0.8\textwidth]{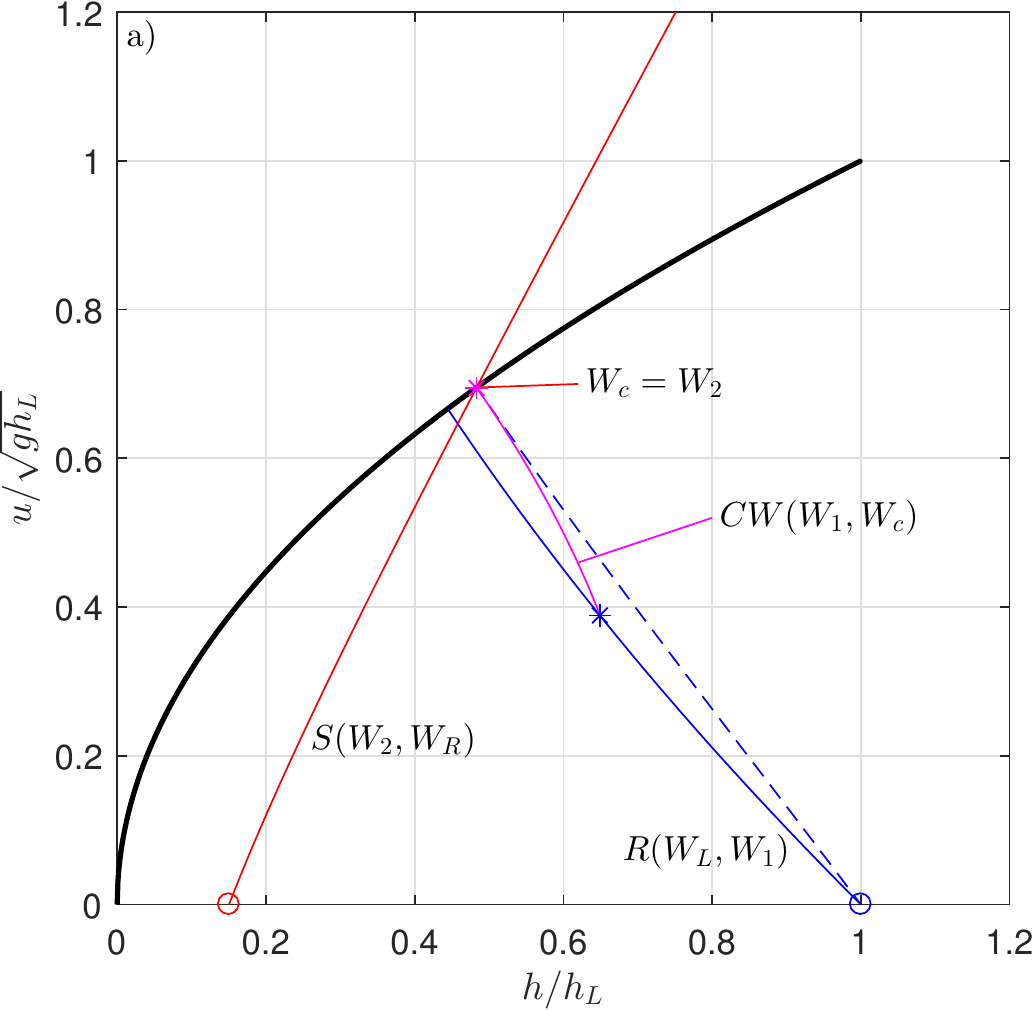}

\rule{0mm}{6mm}

\includegraphics[width=1.0\textwidth]{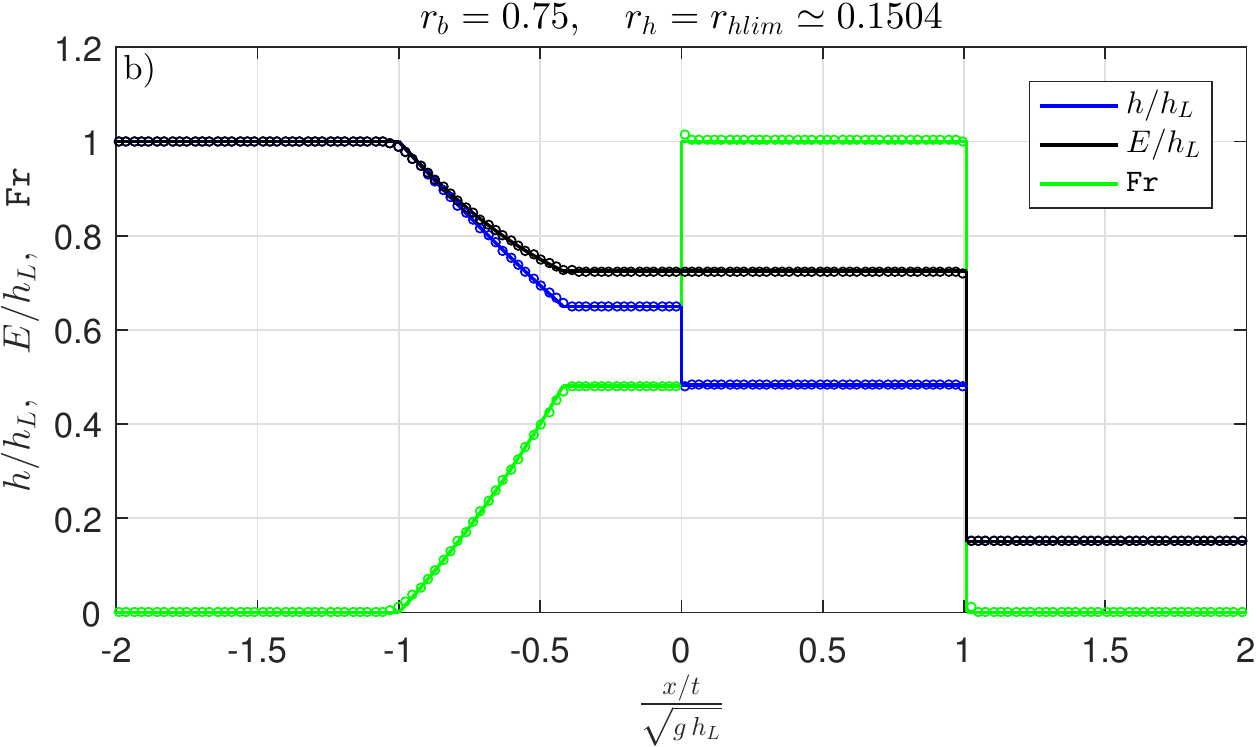}
\end{center}
\caption{Contraction, limit value of $r_h$. a) solution in the phase plane. b) solution in the physical plane. The continuous line is the analytical solution, and the circles represent the numerical solution.}
\label{fig:CLLDR}
\end{figure}

Using the classic analytical solution of the Riemann problem (Fig.~\ref{fig:CLLDR}a), the rarefaction curve $R$ (continuous blue line) is analogous to that of the previous subsections \ref {subsec:CLDR} and \ref {subsec:CSDR}. The subcritical dashed blue line is analogous to that of the previous subsections \ref {subsec:CLDR} and \ref {subsec:CSDR} as well. Starting from the right state, the shock curve (continuous red line) intersects the dashed blue line exactly on the resonance curve so that the red asterisk and magenta asterisk of the previous subsections \ref{subsec:CLDR}, \ref{subsec:CSDR} collapse into a unique point. From this point, the contact wave curve $CW$ (continuous magenta line) is computed, considering a linear variation of the channel width from $b_R$ to $b_L$, up to the intersection with the first $R$ curve (blue asterisk). The blue asterisk and the magenta asterisk identify the two constant states upstream and downstream of the dam, respectively, which appear in the following. The latter also coincides with the critical state. The solution is governed by the following system of equations (moving from upstream to downstream).

The identity of the critical state and the constant state 2, Eq.~\eqref{eq:h2_oneq} is modified as follows:
\begin{equation}\label{eq:h2=hc}
\sqrt{g\, h_c}=\left(h_c-h_R\right) \sqrt{\frac{1}{2}\,g\left(\frac{1}{h_c}+\frac{1}{h_R} \right)}
\end{equation}

Let be $\rho_c={h_c}/{h_R}$; the symbol $\rho$ is used instead of $r$ solely to emphasize that the ratio is obtained by scaling with respect to $h_R$ instead of $h_L$.
From Eq.~\eqref{eq:h2=hc}, a third-degree equation is obtained in the following form:
\begin{equation}\label{eq:rsmall}
\rho_c^3 - 3\,\rho_c^2 - \rho_c + 1 = 0
\end{equation}
Eq.~\eqref{eq:rsmall} has the three solutions $(\rho_c^{(1)} \simeq 3.2143, \, \rho_c^{(2)} \simeq 0.4608, \, \rho_c^{(3)}\simeq -0.6751)$.
Only the first solution has physical validity because $\rho_c^{(1)} > 1$. Moreover, $\rho_{clim}=\rho_c^{(1)}$ is the limit ratio between the critical depth at $x=0^{+}$ and the initial right depth, dividing the \emph{large} depth solution from the \emph{small} depth solution.

Given:
\begin{equation}\label{eq:alpha}
\alpha = \arccos \left(\frac{3 \, \sqrt{3}}{8}\right)
\end{equation}
the exact $\rho_{c}^{(1)}$ value is given by:
\begin{equation}\label{eq:rho_lim}
\rho_{clim} =  \left(\frac{h_c}{h_R}\right)_{lim} = 1 + \frac{4}{\sqrt{3}} \, \cos \left(\frac{\alpha}{3} \right)
\end{equation}
This limit value corresponds to the critical stage at $x=0^{+}$ and to the limit conditions: $h_2=h_c$, $u_2=u_c$, and $x_{20}=0$.

The next step is to find the limit condition in the fundamental plane governing the problem, ($r_b$, $r_h$). From Eqs.~\eqref{eq:R vincM} and \eqref{eq:R vincm}, with $r_c=\rho_c \, r_h$, the lower and upper limits for the depth ratio in the limit case are:
\begin{subequations}
\label{eq:rhminmax}
\begin{align}
r_{hlim} &\ge \frac{4/9}{\rho_{clim}}\simeq 0.1383
\label{eq:rhmin}\\
r_{hlim} &< \frac{2/3}{\rho_{clim}}\simeq 0.2074
\label{eq:rhmax}
\end{align}
\end{subequations}

The final conclusion is that, in the range of $0.1383 \le r_{hlim} < 0.2074$, it is possible to identify the limit curve between the \emph{large} and \emph{small} depth ratios, simply moving the $r_{clim}$ parameter within the range $[4/9; \, 2/3)$ and computing the corresponding values of $r_{1lim}$, from Eq.~\eqref{eq:phi2}; of $r_b$, from Eq.~\eqref{eq:rb}: it is simple to verify that this result varies in the range $(0; \, 1]$, as is appropriate for the case of contraction. The limit curve is drawn in the final diagram, Fig.~\ref{fig:CLIMR}.

For any possible case in practice, if the point ($r_b$, $r_h$) lies over the limit curve, the solution is a \emph{large} depth ratio solution (green area); if the point lies under the same curve, the solution is a \emph{small} depth ratio solution (blue area).
The described procedure allows us to find the couple ($r_b$, $r_{hlim}$) identifying the limit conditions, starting from $r_{clim}$ (and from the corresponding $r_{hlim})$ and then finding the corresponding  $r_{b}$.

If we consider a fixed $r_{b}$, to obtain the corresponding $r_{hlim}$, we have to consider Eq.~\eqref{eq:phi2} and Eq.~\eqref{eq:rb}, so that $r_{b}$ can be expressed as a function $r_b = f(r_{clim})$ only:
\begin{equation}\label{eq:r_bgfunct}
r_{b} = \frac{2}{r_{clim}^{3/2}} \, \left( \frac{2}{3} + \frac{\sqrt{2}}{2}\,\sqrt{r_{clim} -  \frac{4}{9}\,}\right)^2 \, \left[ 1-\left( \frac{2}{3} + \frac{\sqrt{2}}{2}\,\sqrt{r_{clim} -  \frac{4}{9}\,}\right)\right]
\end{equation}

Equation \eqref{eq:r_bgfunct} must be inverted using a Newton method. A linearly interpolated value is a good option for the initial guess value.
In such a way, the limit curve can be simply found as follows:
\begin{equation}\label{eq:g_1}
 r_{clim}=f^{-1}(r_b); \quad  r_{hlim}=\frac{r_{clim}}{\rho_{clim}}
\end{equation}
where $f^{-1}$ is the inverse function of $f$ and $\rho_{clim}$ is given by Eq.~\eqref{eq:rho_lim}.

The typical solution is depicted (continuous line) in the physical plane in Fig.~\ref{fig:CLLDR}b, together with the corresponding numerical solution (circles).

\section{Analytical solution for channel expansion}
\label{sec:exp}
In the case of $b_R>b_L$, it is necessary to consider four different kinds of solutions that occur for \emph{large}, \emph{intermediate}, \emph{small} and \emph{very small} values of the downstream/upstream depth ratio $r_h$, respectively.

A \emph{first upper limit} value $\overline{r}_{hlim}$ of the depth ratio divides the first type of solution from the second type. A \emph{second upper limit} value $r_{hlim}$ of the depth ratio divides the second type of solution from the third type. A \emph{lower limit} value $r_{hlim}^{*}$ of the depth ratio divides the third type of solution from the fourth type.
All three limits depend on the downstream/upstream width ratio $r_b$, as can be expected on a physical basis; therefore, they correspond to three curves in the ($r_b$, $r_h$) plane.

\subsection{Expansion, large depth ratio}
\label{subsec:ELDR}
This solution occurs when the initial right depth is large enough to obtain a subcritical flow everywhere. Looking at the problem from upstream to downstream, after the constant still-water state, an upstream rarefaction exists, whose head moves upstream with a negative celerity, and then a sequence of two constant states (instead of the only one for the constant width problem) divided by a contact wave at the dam position, a downstream moving shock with a positive celerity, and a still-water downstream state occur. The downstream boundary of the second constant state is a moving shock. In the case of expansion, at the dam position $x=0$, the contact wave corresponds to an increase in depth and a decrease in velocity.

\begin{figure}
\begin{center}
\includegraphics[width=0.8\textwidth]{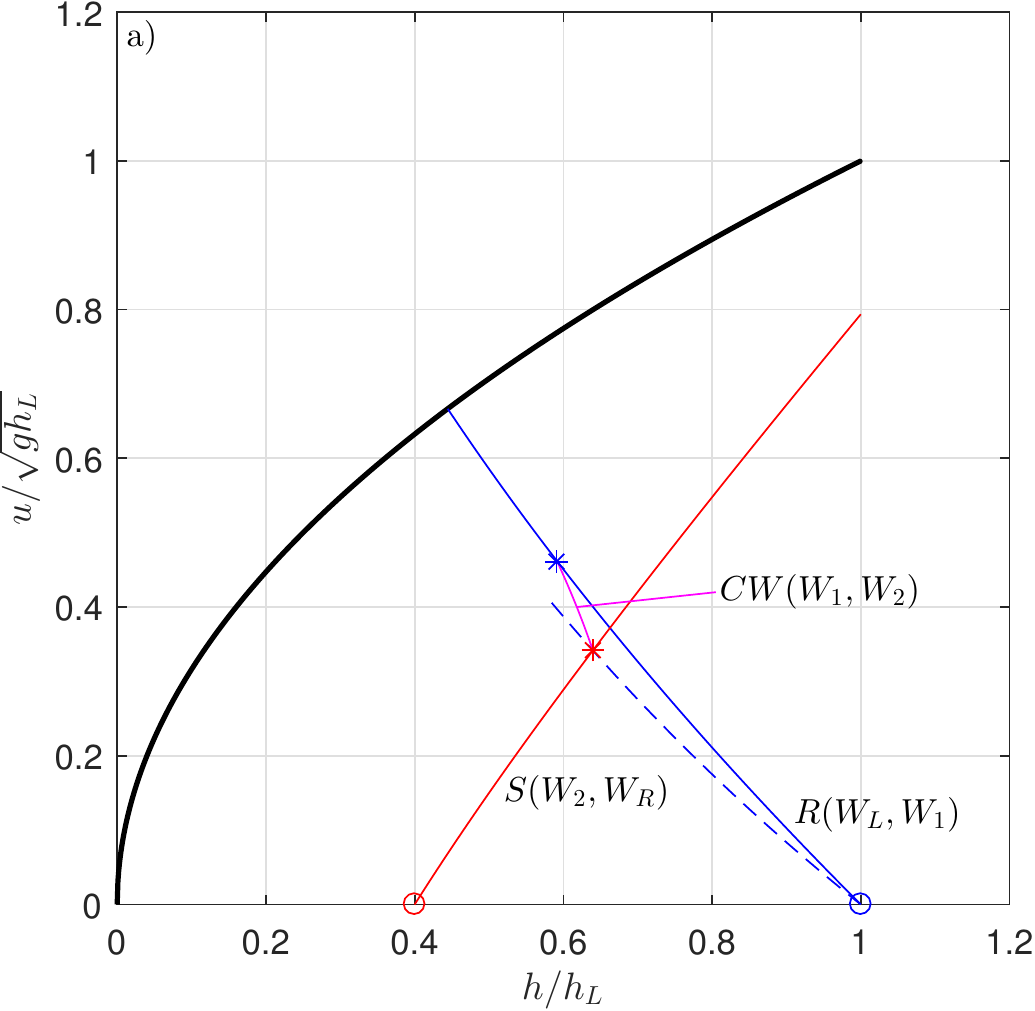}

\rule{0mm}{6mm}

\includegraphics[width=1.0\textwidth]{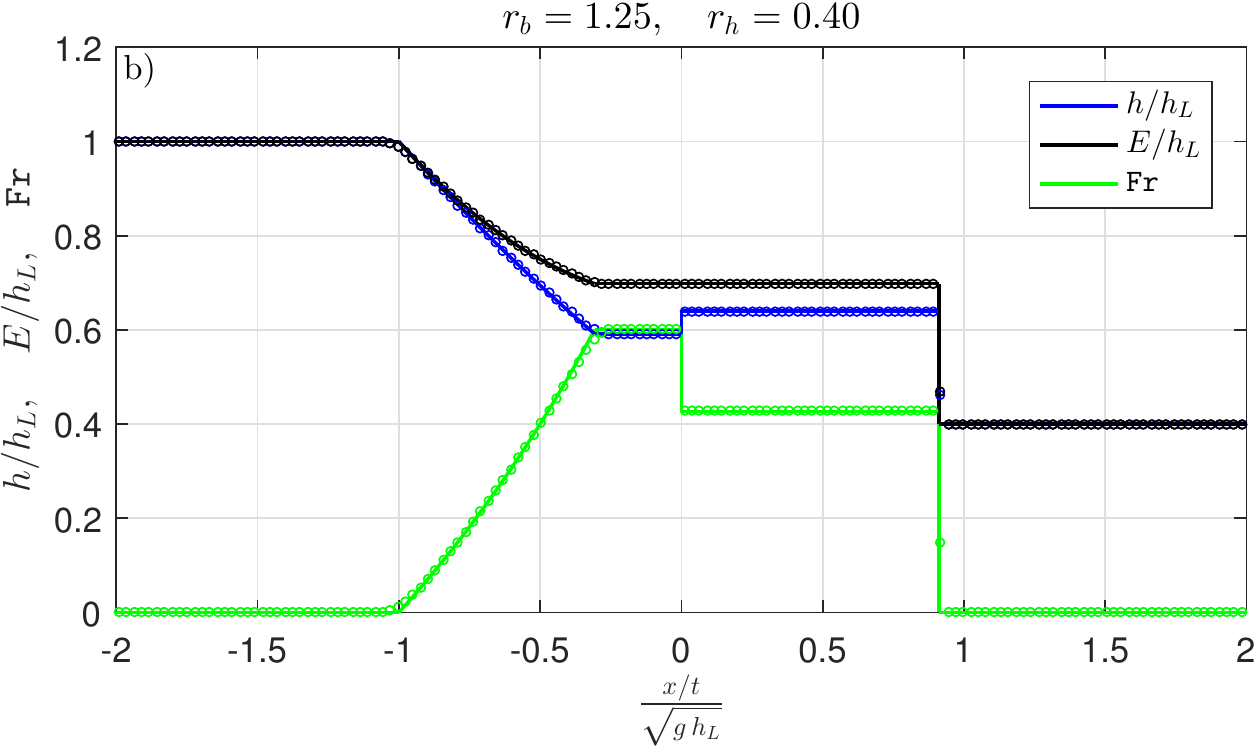}
\end{center}
\caption{Expansion, large $r_h$. a) solution in the phase plane. b) solution in the physical plane. The continuous line is the analytical solution, and the circles represent the numerical solution.}
\label{fig:ELDR}
\end{figure}

Using the classic analytical solution of the Riemann problem (Fig.~\ref{fig:ELDR}a), the procedure is analogous to that of subsection \ref {subsec:CLDR}: a rarefaction curve $R$ (continuous blue line) is computed, starting from the left state up to the intersection with the critical resonance curve (thick black line). A corresponding subcritical curve for the different width is computed, and the corresponding dashed blue is drawn. Starting from the right state, a shock curve $S$ (continuous red line) is computed, which intersects the dashed blue line at a point (red asterisk). From this point, the contact wave curve $CW$ (continuous magenta line) is computed, which intersects the first $R$ curve (blue asterisk). The blue asterisk and the red asterisk are the two constant states upstream and downstream of the dam, respectively.

The computations are strictly similar to those described in subsection \ref{subsec:CLDR}. The system of Eqs.~\eqref{eq:UPRAR1}, \eqref{eq:MCD}, \eqref{eq:ECD}, \eqref{eq:RHx2} in the four unknowns $h_1, u_1, h_2, u_2$ is solved using the Newton-Raphson method. Then, the downstream shock celerity $c_2$ is found using Eq.~\eqref{eq:c2}, so that a complete framework is available.

The typical solution is depicted (continuous line) in the physical plane Fig.~\ref{fig:ELDR}b, together with the corresponding numerical solution (circles).

\subsection{Expansion, intermediate depth ratio}
\label{subsec:EIDR}
This solution occurs when the initial right depth is large enough to obtain a subcritical flow upstream and downstream the dam, but a stationary shock occurs \emph{at} the dam. Intentionally, the example presented here is obtained for the same depth ratio ($r_h=0.40$) of the previous subsection \ref{subsec:ELDR}, but using a larger width ratio ($r_b=2.75$), to stress the relevance of the latter. The upstream rarefaction reaches the critical (sonic) condition at the dam position. Supposing that the channel width at the dam is variable according a prescribed path $b_L \leq b(s) \leq b_R$,  $0 \leq s \leq 1$ (see also section \ref{sec:NMCW} for a proper discussion), a special value of the width is found, $b_*=b(s_*)$, where the shock takes place. The critical flow just upstream the dam position and the subcritical flow just downstream the dam have the same specific energy of the supercritical and subcritical conjugate states of the jump, respectively, characterizing two contact waves, occurring upstream and downstream the jump itself. The constant state in the downstream segment, the downstream moving shock and the still-water right  state are similar to the configuration described in subsection \ref{subsec:ELDR}.

\begin{figure}
\begin{center}
\includegraphics[width=0.8\textwidth]{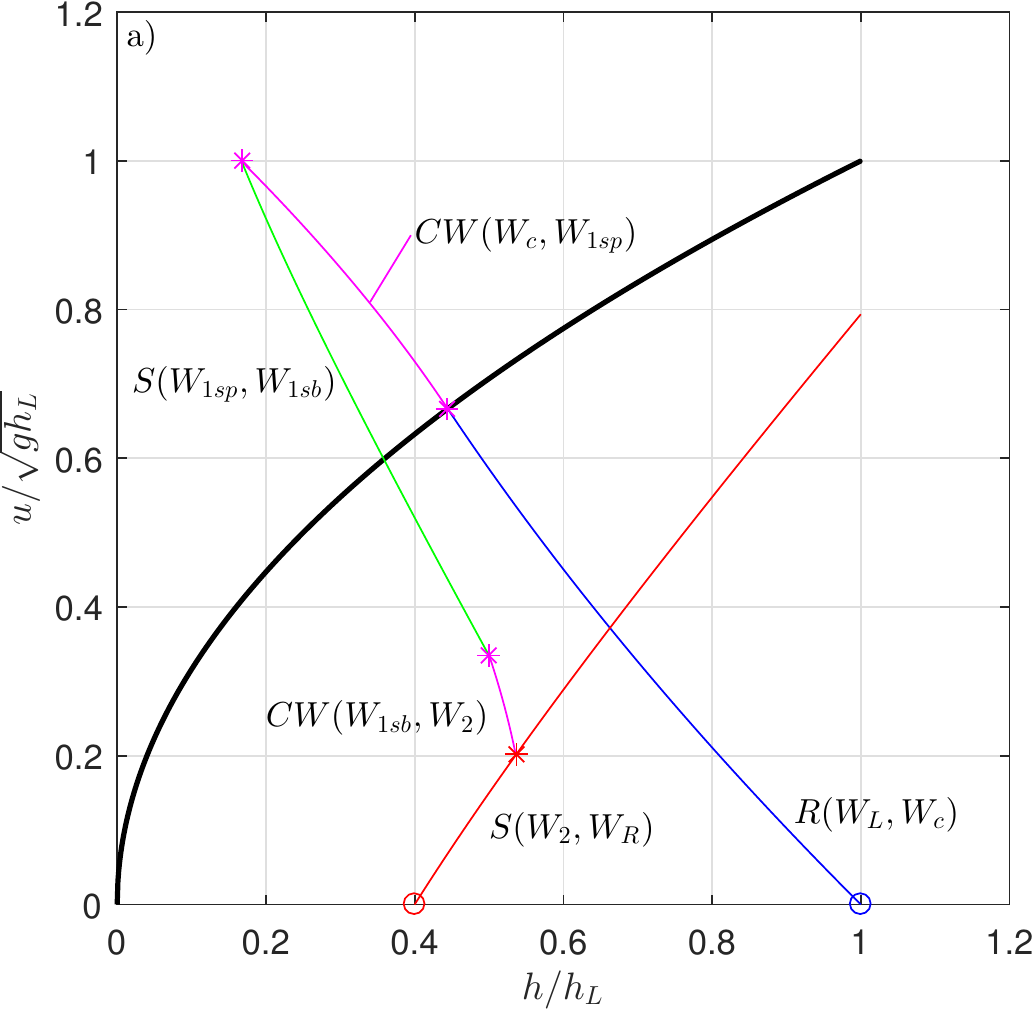}

\rule{0mm}{6mm}

\includegraphics[width=1.0\textwidth]{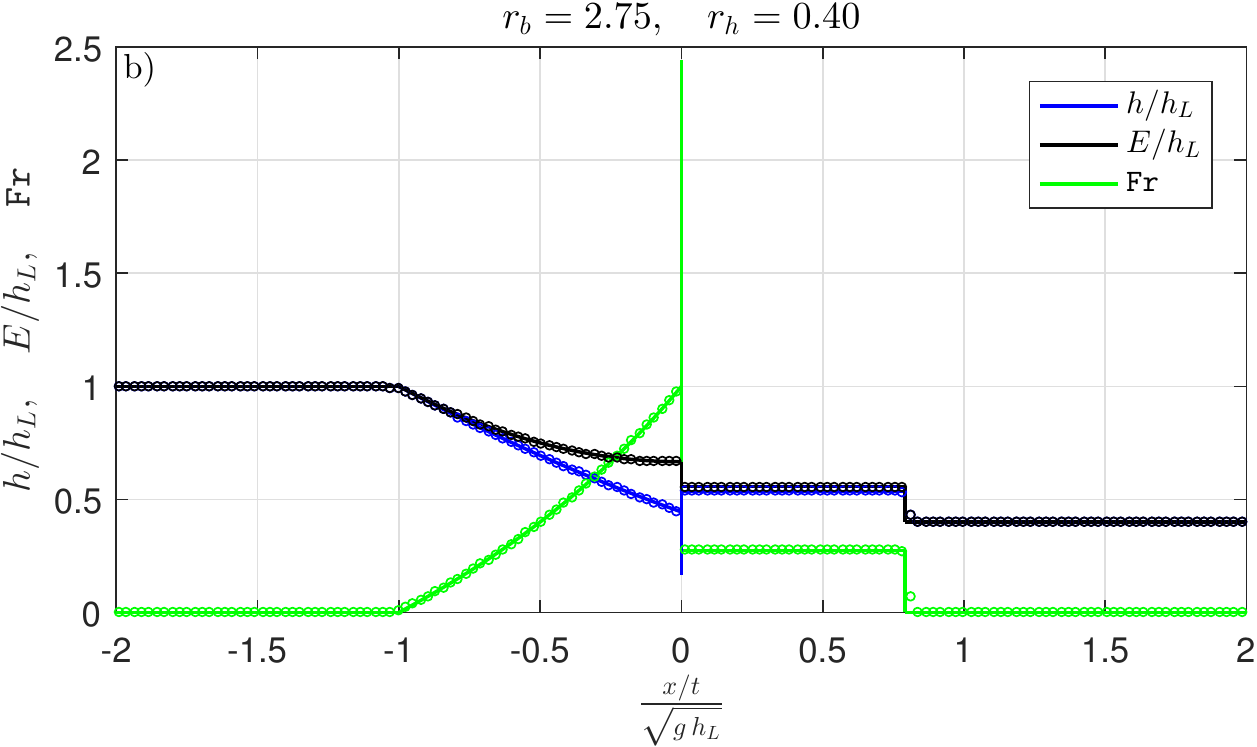}
\end{center}
\caption{Expansion, intermediate $r_h$. a) solution in the phase plane. b) solution in the physical plane. The continuous line is the analytical solution, and the circles represent the numerical solution.}
\label{fig:EIDR}
\end{figure}

Using the classic analytical solution of the Riemann problem (Fig.~\ref{fig:EIDR}a), the procedure identifies a rarefaction curve $R$ (continuous blue line), starting from the left state, up to the intersection with the critical resonance curve (thick black line), where the critical state is indicated by a magenta asterisk. A contact wave (the width varying from $b_L$ to $b_*$) is drawn, up to  the supercritical depth of the shock (magenta asterisk). The green line represents the shock (at $b=b_*$), joining the subcritical depth of the shock (magenta asterisk). A further contact wave (the width varying from $b_*$ to $b_R$) locates the intersection with the shock curve  (continuous red line) originated from the right state (red asterisk).

The solution is governed by the following system of equations (moving from upstream to downstream).

The upstream rarefaction gives a simple one-equation, one-unknown relationship between the left state and the critical state, that is:
\begin{equation}\label{eq:uprarefaction2}
u_L+2\, \sqrt{g\,h_L}=u_c+2\, \sqrt{g\,h_c}=3\sqrt{g\,h_c}
\end{equation}
so that:
\begin{equation}\label{eq:simple2}
h_c=\frac{4}{9}\, h_L; \quad u_c=\frac{2}{3} \, \sqrt{g\,h_L}
\end{equation}
The mass conservation and energy conservation at $x=0$ (from $b=b_L$ to $b=b_*$) give:
\begin{equation}\label{eq:CWSP1}
u_c\, b_L \, h_c = u_{1sp} \, b_* \, h_{1sp}
\end{equation}
\begin{equation}\label{eq:CWSP2}
\frac{3}{2}h_c = h_{1sp} +\frac{u_{1sp}^2}{2\, g}
\end{equation}
where the subscripts \emph{sp} and \emph{sb} (in the following) mean supercritical and subcritical, respectively.
 
The Rankine-Hugoniot condition for the stationary shock and the mass conservation across the shock at $x=0$,  $b=b_*$ give:
\begin{equation}\label{eq:RH00}
u_{1sp} = h_{1sb} \, \sqrt{\frac{1}{2}\,g\left(\frac{1}{h_{1sp}}+\frac{1}{h_{1sb}} \right)}
\end{equation}
\begin{equation}\label{eq:mass0}
u_{1sp} \, h_{1sp} = u_{1sb} \, h_{1sb}
\end{equation}

The mass conservation and energy conservation at $x=0$ (from $b=b_*$ to $b=b_R$) give:
\begin{equation}\label{eq:CWSB1}
u_{1sb} \, b_* \, h_{1sb} = u_2\, b_R \, h_2 
\end{equation}
\begin{equation}\label{eq:CWSB2}
h_{1sb} +\frac{u_{1sb}^2}{2\, g} = h_{2} +\frac{u_{2}^2}{2\, g}
\end{equation}

The Rankine-Hugoniot condition for the downstream moving shock at $x=x_2=c_2 \, t$ gives Eq.~\eqref{eq:RHx2}, whereas Eq.~\eqref{eq:c2} gives the shock celerity.

The system of Eqs.~\eqref{eq:CWSP1}, \eqref{eq:CWSP2}, \eqref{eq:RH00}, \eqref{eq:mass0}, \eqref{eq:CWSB1}, \eqref{eq:CWSB2}, \eqref{eq:RHx2} in the seven unknowns $b_*, h_{1sp}, u_{1sp}, h_{1sb}, u_{1sb}, h_2, u_2$ is solved using the Newton-Raphson method.

Once this computation is completed, the final solution is the following.

A constant left state upstream:
\begin{equation}\label{eq:CLSb}
\left\{ \begin{array}{l} h = h_L \\ u = 0 \end{array} \right . \quad \textrm{for} \quad \frac{x}{t} \leq \frac{x_L}{t}=-\left(\sqrt{g\,h_L} \right)   
\end{equation}

A rarefaction, associated with the first (negative) eigenvalue:
\begin{equation}\label{eq:UPRARb}
u_L+2\, \sqrt{g\,h_L}=u+2\, \sqrt{g\,h}=3\sqrt{g\,h_c} \quad \textrm{for} \quad \frac{x_L}{t} \leq \frac{x}{t} \leq 0^{-}   
\end{equation}

The critical state $(h_c, u_c)$ at $x=0^{-}$; conventionally, the supercritical branch of the contact contact wave for $0^{-} < x <0$, the stationary shock, joining $(h_{1sp}, u_{1sp})$ with $(h_{1sb}, u_{1sb})$, at $x=0$,  the supercritical branch of the contact contact wave for $0< x <0^{+}$; the constant state $(h_2, u_2)$ begins at  $x=0^{+}$.

A constant state, just downstream of the initial dam position:
\begin{equation}\label{eq:CSDD}
\left\{ \begin{array}{l} h = h_2 \\ u = u_2 \end{array} \right . \quad \textrm{for} \quad 0^{+} \leq \frac{x}{t} \leq \frac{x_{2}}{t}=c_2   
\end{equation}

A constant state, just downstream of the final shock, that corresponds to the downstream initial state:
\begin{equation}\label{eq:CRSb}
\left\{ \begin{array}{l} h = h_R \\ u = 0 \end{array} \right . \quad \textrm{for} \quad \frac{x}{t} \geq \frac{x_{2}}{t}   
\end{equation}

The framework for the expansion, \emph{intermediate} depth ratio is now completed.
The typical solution is depicted (continuous line) in the physical plane in Fig.~\ref{fig:EIDR}b, together with the corresponding numerical solution (circles).

\subsection{Expansion, first upper limit depth ratio}
\label{subsec:EIDRLIM}
This subsection analytically finds the limit curve dividing a \emph{large} depth ratio from an \emph{intermediate} depth ratio in the case of expansion to identify the range of existence of both. Inspecting the structures of the flow field described in the previous subsections \ref{subsec:ELDR} and \ref{subsec:EIDR}, we recognize the limit as occurring when $b_*=b_L$, the stationary shock disappears and consequently $E_c=E_2$, that is, only a contact wave occurs at the dam position.

A further limit curve will be studied in the following section \ref{subsec:EULDR}, which corresponds to $b_*=b_R$.

\begin{figure}
\begin{center}
\includegraphics[width=0.8\textwidth]{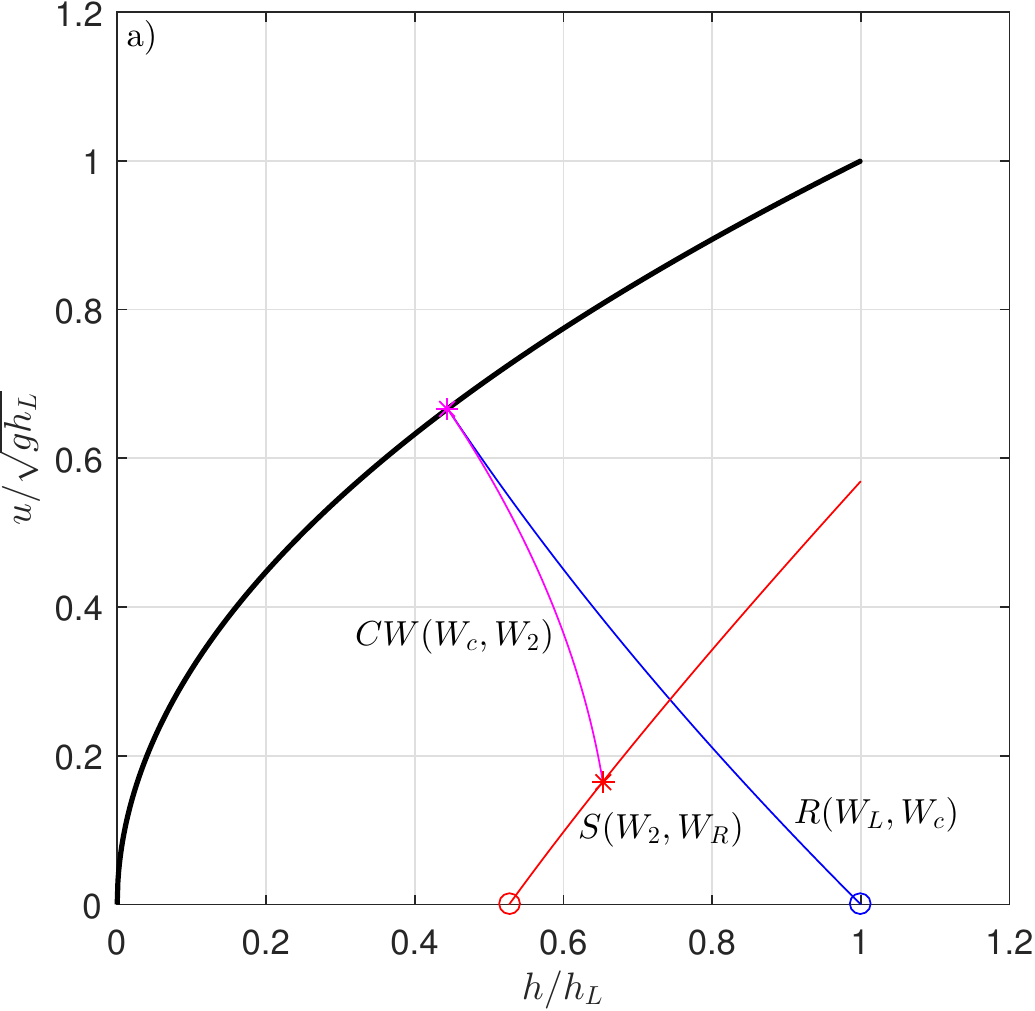}

\rule{0mm}{6mm}

\includegraphics[width=1.0\textwidth]{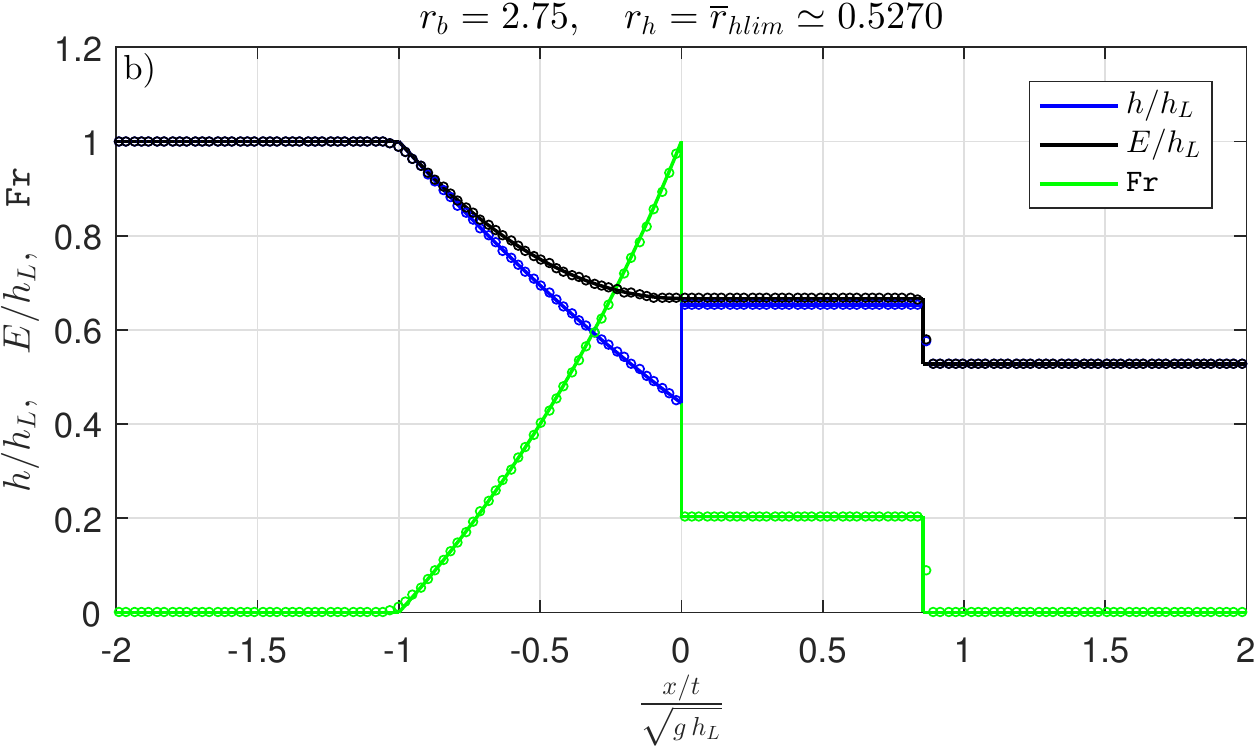}
\end{center}
\caption{Expansion, first upper limit depth ratio. a) solution in the phase plane. b) solution in the physical plane. The continuous line is the analytical solution, and the circles represent the numerical solution.}
\label{fig:EIDRLIM}
\end{figure}

Using the classic analytical solution of the Riemann problem (Fig.~\ref{fig:EIDRLIM}a), the reasoning is simpler than that of the previous subsection \ref{subsec:EIDR}. The rarefaction curve $R$ (continuous blue line) is drawn up to the resonance curve. The magenta asterisk identifies the critical point at $0^{-}$. From this point, the contact wave curve $CW$ (continuous magenta line) is computed up to the intersection with the right shock curve (continuous red line) at a point (red asterisk), which identifies the ($h_2$, $u_2$) condition.

At the dam position, the mass conservation and energy conservation give:
\begin{equation}\label{eq:fupm}
u_c\, b_L \, h_c = u_2 \, b_R \, h_2
\end{equation}
\begin{equation}\label{eq:fupe}
\frac{3}{2}h_c = h_2 +\frac{u_2^2}{2\, g}
\end{equation}

Using Eq.~\eqref{eq:fupm} to derive $u_2$, after some algebra Eq.~\eqref{eq:fupe} gives the following third-degree equation:
\begin{equation}\label{eq:3rdfup}
\left(\frac{h_c}{h_2}\right)^3 - 3 \, r_b^2  \left(\frac{h_c}{h_2}\right) + 2 \, r_b^2 = 0
\end{equation}

Let the following apply:
\begin{equation}\label{eq:gam_rb}
\gamma = \arccos \left(-\frac{1}{r_b^2} \right)
\end{equation}

Being $r_b>1$, $\gamma$ ranges from $\pi$ to $\pi/2$ for $r_b$ ranging in the interval $(1; \infty)$; in the same range, Eq.~\eqref{eq:3rdfup} has the following three solutions:
\begin{subequations}
\begin{align}
\left(\frac{h_c}{h_2}\right)^{(1)} &= 2 \, r_b \, \cos \left(\frac{\gamma}{3}\right) &\Rightarrow& & 1<& \left(\frac{h_c}{h_2}\right)^{(1)} < \infty
\label{eq:solu1}\\
\left(\frac{h_c}{h_2}\right)^{(2)} &= 2 \, r_b \, \cos \left(\frac{\gamma}{3}+\frac{2\pi}{3}\right) &\Rightarrow& & -2>& \left(\frac{h_c}{h_2}\right)^{(2)} > -\infty
\label{eq:solu2}\\
\left(\frac{h_c}{h_2}\right)^{(3)} &= 2 \, r_b \, \cos \left( \frac{\gamma}{3}-\frac{2\pi}{3}\right) & \Rightarrow& & 1>& \left(\frac{h_c}{h_2}\right)^{(3)} >2/3
\label{eq:solu3}
\end{align}\label{eq:solu13}
\end{subequations}

The third solution \eqref{eq:solu3} is chosen as the only one with physical validity, because the state ($h_2$, $u_2$) must be subcritical.

Consequently, the nondimensional depth in $x=0^{+}$ under the \emph{first upper limit condition} is computed as:
\begin{equation}\label{eq:h2suhLL}
\left(\frac{h_2}{h_L}\right)_{lim} = \left(\frac{h_2}{h_c}\right)^{(3)} \left(\frac{h_c}{h_L}\right) = \frac{4}{9} \left[2 \, r_b \, \cos \left( \frac{\gamma}{3}-\frac{2\pi}{3}\right)\right]^{-1}
\end{equation}
and the corresponding Froude number is:
\begin{equation}\label{eq:Fr_2_uL}
\Fr_2=\frac{u_2}{\sqrt{g\,h_2}} = \sqrt{3\, \frac{h_c}{h_2}-2}
\end{equation}

The last condition to be satisfied is the Rankine-Hugoniot relation on the final shock, that is \eqref{eq:RHx2}. Such equation is made nondimensional as:
\begin{equation}\label{eq:Fr2_uLL}
\Fr_2=\frac{u_2}{\sqrt{g \, h_2}} = \left( 1 - \frac{h_R}{h_2}  \right)  \sqrt{\frac{1}{2}\left(1+\frac{h_2}{h_R} \right)}
\end{equation}
Using $({h_R}/{h_2})$ as an independent variable, Eq.~\eqref{eq:Fr2_uLL} can be written in the form of a third-degree polynomial equation:
\begin{equation}\label{eq:psi3rd}
\left(\frac{h_R}{h_2}\right)^3 - \left(\frac{h_R}{h_2}\right)^2 - \left(1+2 \, \Fr_2^2 \right) \, \left(\frac{h_R}{h_2}\right) + 1 = 0
\end{equation}
Such an equation has three solutions, but only one is in the proper range $0 < ({h_R}/{h_2}) < 1$, that is, the following (with $\left(1+2 \, \Fr_2^2 \right)>1$, one different solution is greater than 1, and the last one is negative):
\begin{equation}\label{eq:hR2uL}
\left(\frac{h_R}{h_2}\right)= \frac{1}{3} - \frac{2}{3} \sqrt{3\left(1+2 \, \Fr_2^2 \right)+1} \, \cos \left( \frac{\theta}{3} + \frac{\pi}{3} \right)
\end{equation}
where:
\begin{equation}\label{eq:theta}
\theta = \arccos \left( \frac{9\left(1+2 \, \Fr_2^2 \right)-25}{2 \sqrt{\left[3\left(1+2 \, \Fr_2^2 \right)+1 \right]^3}} \right)
\end{equation}

In conclusion, the $\overline{r}_{hlim}$ limit ratio can be found, which divides the \emph{large} depth ratio from the \emph{intermediate} depth ratio in the case of expansion, corresponding to an unique contact wave from the critical state to the constant state downstream the dam, as:
\begin{equation}\label{eq:rh_ulim}
\overline{r}_{hlim}=\left(\frac{h_R}{h_L}\right)_{lim}=\left(\frac{h_R}{h_2}\right) \, \left( \frac{h_2}{h_L} \right)_{lim}
\end{equation}
where $(h_R/h_2)$ is obtained from Eq.~\eqref{eq:hR2uL} and $(h_2/h_L)_{lim}$ is obtained from Eq.~\eqref{eq:h2suhLL}.

Referring to Fig.~\ref{fig:CLIMR}, if the point $(r_b, r_h)$ lies over the limit curve, the solution is a \emph{large} depth ratio solution (yellow area); if the point lies under the same curve and over the second upper limit curve, the solution is an \emph{intermediate} depth ratio solution (magenta area).

The typical solution is depicted (continuous line) in the physical plane in Fig.~\ref{fig:EIDRLIM}b, together with the corresponding numerical solution (circles).

\subsection{Expansion, small depth ratio}
\label{subsec:ESDR}
A \emph{small} downstream/upstream depth ratio, $r_h$, is considered; it produces a supercritical solution downstream of the dam position. The upstream rarefaction reaches the initial dam position, and the critical state occurs for $x=0^{-}$. The solution is resonant because the first eigenvalue is zero at this point and the problem loses strict hyperbolicity. At $x=0^{+}$, the constant state $h_1$, $u_1$ is supercritical, as demonstrated in the following, and a further (with respect to the downstream shock) shock occurs, moving downstream. Such a shock is followed downstream by a constant state $h_2$, $u_2$ and the downstream final shock, moving downstream and dividing the constant state from the final still-water state $h_R$, $u_R=0$.

\begin{figure}
\begin{center}
\includegraphics[width=0.8\textwidth]{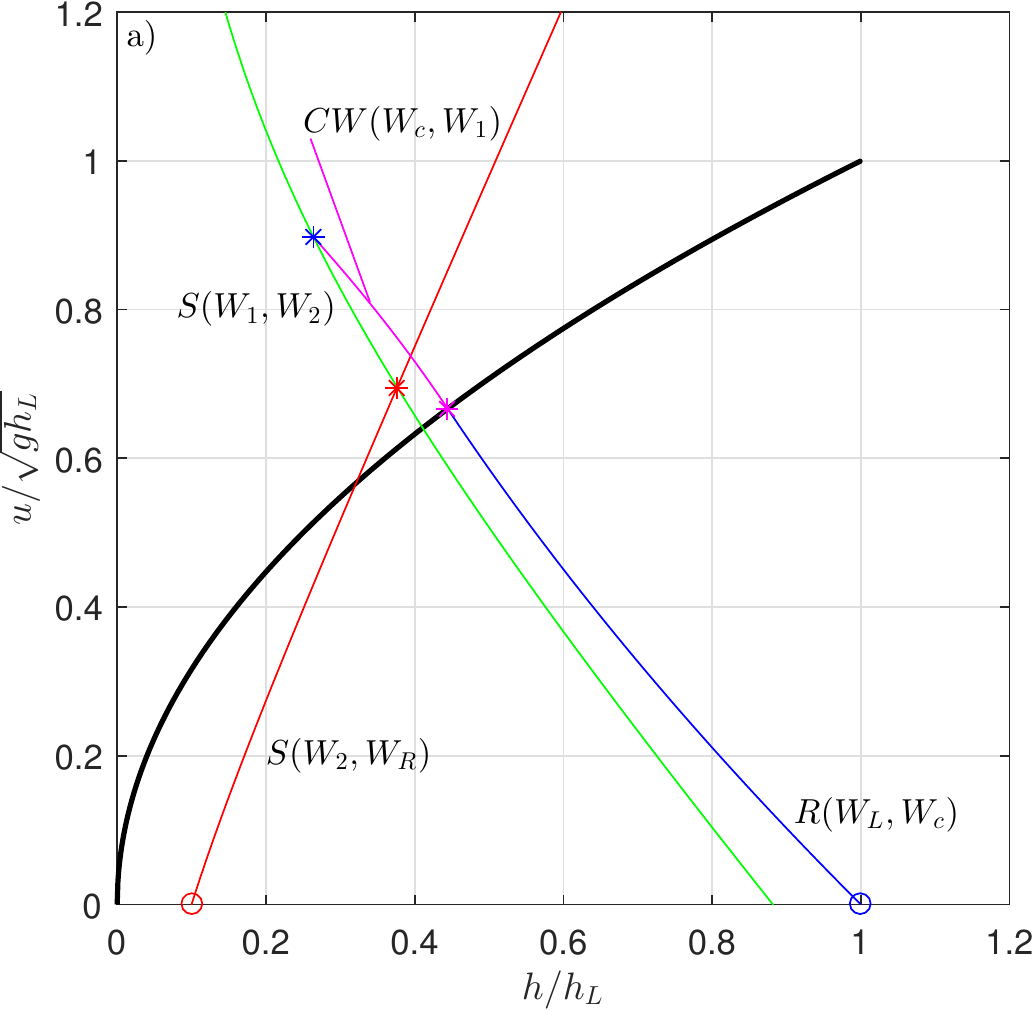}

\rule{0mm}{6mm}

\includegraphics[width=1.0\textwidth]{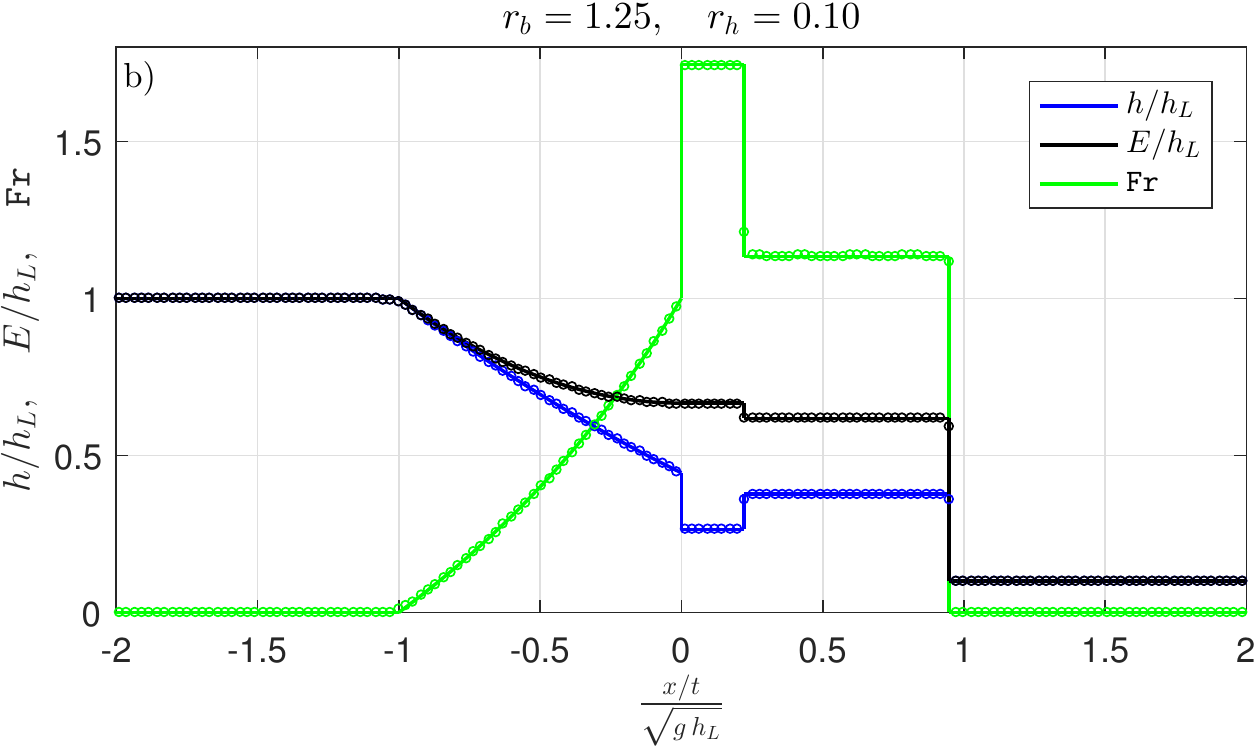}
\end{center}
\caption{Expansion, small $r_h$. a) solution in the phase plane. b) solution in the physical plane. The continuous line is the analytical solution, and the circles represent the numerical solution.}
\label{fig:ESDR}
\end{figure}

Using the classic analytical solution of the Riemann problem (Fig.~\ref{fig:ESDR}a), the rarefaction curve $R$ (continuous blue line) is drawn up to the resonance curve. The magenta asterisk identifies the critical point at $0^{-}$. From this point, the contact wave curve $CW$ (continuous magenta line) is computed, using a linear path for the width, conserving total discharge and specific energy up to $b=b_R$ (blue asterisk), obtaining the ($h_1$, $u_1$) condition. From this point, the shock curve (continuous green line) is drawn, which intersects the right shock curve (continuous red line) at a point (red asterisk), which identifies the ($h_2$, $u_2$) condition. The blue asterisk and the red asterisk identify the two constant states, both downstream of the dam, divided by a downstream moving shock.

The solution is governed by the following system of equations (moving from upstream to downstream).

Eq~\eqref{eq:simple2} still holds at $x=0^{-}$.

The mass conservation and energy conservation at $x=0$ give:
\begin{equation}\label{eq:downconst1}
u_c\, b_L \, h_c = u_1 \, b_R \, h_1
\end{equation}
\begin{equation}\label{eq:downconst2}
\frac{3}{2}h_c = h_1 +\frac{u_1^2}{2\, g}
\end{equation}

Using Eq.~\eqref{eq:downconst1} to derive $u_1$, Eq.~\eqref{eq:downconst2} gives:
\begin{equation}\label{eq:3rd eqn}
\frac{h_1}{h_c} + \frac{1}{2\, r_b^2}\left( \frac{h_c}{h_1} \right)^2 - \frac{3}{2} = 0
\end{equation}

Eq.~\eqref{eq:3rd eqn} corresponds to the following third-degree equation:
\begin{equation}\label{eq:3rd_h1suhc}
\left(\frac{h_1}{h_c}\right)^3 - \frac{3}{2} \left(\frac{h_1}{h_c}\right)^2 + \frac{1}{2\, r_b^2}  = 0
\end{equation}

Let the following apply:
\begin{equation}\label{eq:beta_rb}
\beta = \arccos \left(1 - \frac{2}{r_b^2} \right)
\end{equation}

Remembering that $r_b>1$, we find that $\beta$ ranges from $\pi$ to $0$ for $r_b$ ranging in the interval $(1; \infty)$; in the same range, Eq.~\eqref{eq:3rd_h1suhc} has the following three solutions:
\begin{subequations}
\begin{align}
\left(\frac{h_1}{h_c}\right)^{(1)} &= \frac{1}{2} - \cos \left( \frac{\beta}{3} + \frac{\pi}{3} \right) &\Rightarrow& & 1>& \left(\frac{h_1}{h_c}\right)^{(1)} >0
\label{eq:soluz1}\\
\left(\frac{h_1}{h_c}\right)^{(2)} &= \frac{1}{2} - \cos \left( \frac{\beta}{3} - \frac{\pi}{3} \right) &\Rightarrow& & -1/2<& \left(\frac{h_1}{h_c}\right)^{(2)} <0
\label{eq:soluz2}\\
\left(\frac{h_1}{h_c}\right)^{(3)} &= \frac{1}{2} + \cos \left( \frac{\beta}{3}\right) & \Rightarrow& & 1<& \left(\frac{h_1}{h_c}\right)^{(3)} <3/2
\label{eq:soluz3}
\end{align}\label{eq:soluz13}
\end{subequations}

The first solution \eqref{eq:soluz1} is chosen as the only one with physical validity because the supercritical condition is the only one compatible with a positive celerity of the shock dividing the state ($h_1$, $u_1$) from the state ($h_2$, $u_2$), as shown using Eq.~\eqref{eq:dc1_ad}.

Consequently, the nondimensional depth in $x=0^{+}$ is computed as:
\begin{equation}\label{eq:h1suhL}
\frac{h_1}{h_L} = \frac{4}{9} \left[ \frac{1}{2} - \cos \left( \frac{\beta}{3} + \frac{\pi}{3} \right) \right]
\end{equation}
and the corresponding Froude number is:
\begin{equation}\label{eq:Fr_1_0p}
\Fr_1=\frac{u_1}{\sqrt{g\,h_1}} = \sqrt{3\, \frac{h_c}{h_1}-2}
\end{equation}

The Rankine-Hugoniot condition for the shock dividing the state ($h_1$, $u_1$) from the state ($h_2$, $u_2$) is:
\begin{equation}\label{eq:ds1}
u_2 = u_1 - \left(h_2-h_1\right) \sqrt{\frac{1}{2}\,g\left(\frac{1}{h_2}+\frac{1}{h_1} \right)}
\end{equation}
and the corresponding celerity is:
\begin{equation}\label{eq:dc1}
c_1 = u_1 - h_2\sqrt{\frac{1}{2}\,g\left(\frac{1}{h_2}+\frac{1}{h_1} \right)}
\end{equation}
Making nondimensional \eqref{eq:dc1} by dividing both sides of the equation by $\sqrt{g\,h_1}$, we obtain:
\begin{equation}\label{eq:dc1_ad}
\frac{c_1}{\sqrt{g\,h_1}} = \Fr_1 - \frac{h_2}{h_1}\sqrt{\frac{1}{2}\left(\frac{h_1}{h_2}+1 \right)}
\end{equation}
To obtain a positive celerity in Eq.~\eqref{eq:dc1_ad}, being $h_2>h_1$ as required by Eq.~\eqref{eq:ds1}, $\Fr_1>1$ is necessary, so that the assumption of a supercritical flow immediately downstream of the dam position is demonstrated.

The Rankine-Hugoniot condition for the shock dividing the state ($h_2$, $u_2$) from the state ($h_R$, $u_R$) is still Eq.~\eqref{eq:RHx2}, with the corresponding celerity given by Eq.~\eqref{eq:c2}.

The system of equations \eqref{eq:ds1} and \eqref{eq:RHx2} can be solved using the Newton-Raphson method to obtain $h_2$, $u_2$ values, once $h_1$ and $u_1$ are computed using Eqs.~\eqref{eq:h1suhL} and \eqref{eq:Fr_1_0p}, respectively. Then, $c_1$ and $c_2$ are obtained by Eqs.~\eqref{eq:dc1} and \eqref{eq:c2}, respectively.

Once this computation is completed, the final solution is the following.

A constant left state upstream and a following rarefaction, as described by Eqs.~\eqref{eq:CLSb} and \eqref{eq:UPRARb}.

The critical state $(h_c, u_c)$ at $x=0^{-}$.

In $x=0$, the eigenvalue $u-\sqrt{g \, h}$ changes sign, becoming positive in $x=0^{+}$.

A constant state, just downstream of the initial dam position:
\begin{equation}\label{eq:CSDD}
\left\{ \begin{array}{l} h = h_1 \\ u = u_1 \end{array} \right . \quad \textrm{for} \quad 0^{+} \leq \frac{x}{t} \leq \frac{x_{20}}{t}=c_1   
\end{equation}

A constant state, just downstream of the first shock:
\begin{equation}\label{eq:CSD2}
\left\{ \begin{array}{l} h = h_2 \\ u = u_2 \end{array} \right . \quad \textrm{for} \quad \frac{x_{20}}{t} \leq \frac{x}{t} \leq \frac{x_{2}}{t}=c_2    
\end{equation}

A constant state, just downstream of the final shock, corresponds to the downstream initial state:
\begin{equation}\label{eq:CRSb}
\left\{ \begin{array}{l} h = h_R \\ u = 0 \end{array} \right . \quad \textrm{for} \quad \frac{x}{t} \geq \frac{x_{2}}{t}   
\end{equation}

The framework for the expansion, \emph{small} depth ratio is now completed.
The typical solution is depicted (continuous line) in the physical plane in Fig.~\ref{fig:ESDR}b, together with the corresponding numerical solution (circles).

\subsection{Expansion, second upper limit depth ratio}
\label{subsec:EULDR}
This subsection analytically finds the limit curve dividing an \emph{intermediate} depth ratio from a \emph{small} depth ratio in the case of expansion to identify the range of existence of both. Inspecting the structures of the flow field described in the previous subsections \ref{subsec:EIDR} and \ref{subsec:ESDR}, we recognize the limit as occurring when $c_1=0$, that is, when the first shock is stationary and consequently positioned at the dam position, also corresponding to the special value of the channel width $b_*=b_R$.
A further limit curve will be studied in the following section, which is the lower bound for \emph{small} depth ratios, which consequently are included between two limit curves. Each curve, as apparent from the previous subsections, depends on the downstream/upstream width ratio $r_b$, as can be expected on physical bases.

\begin{figure}
\begin{center}
\includegraphics[width=0.8\textwidth]{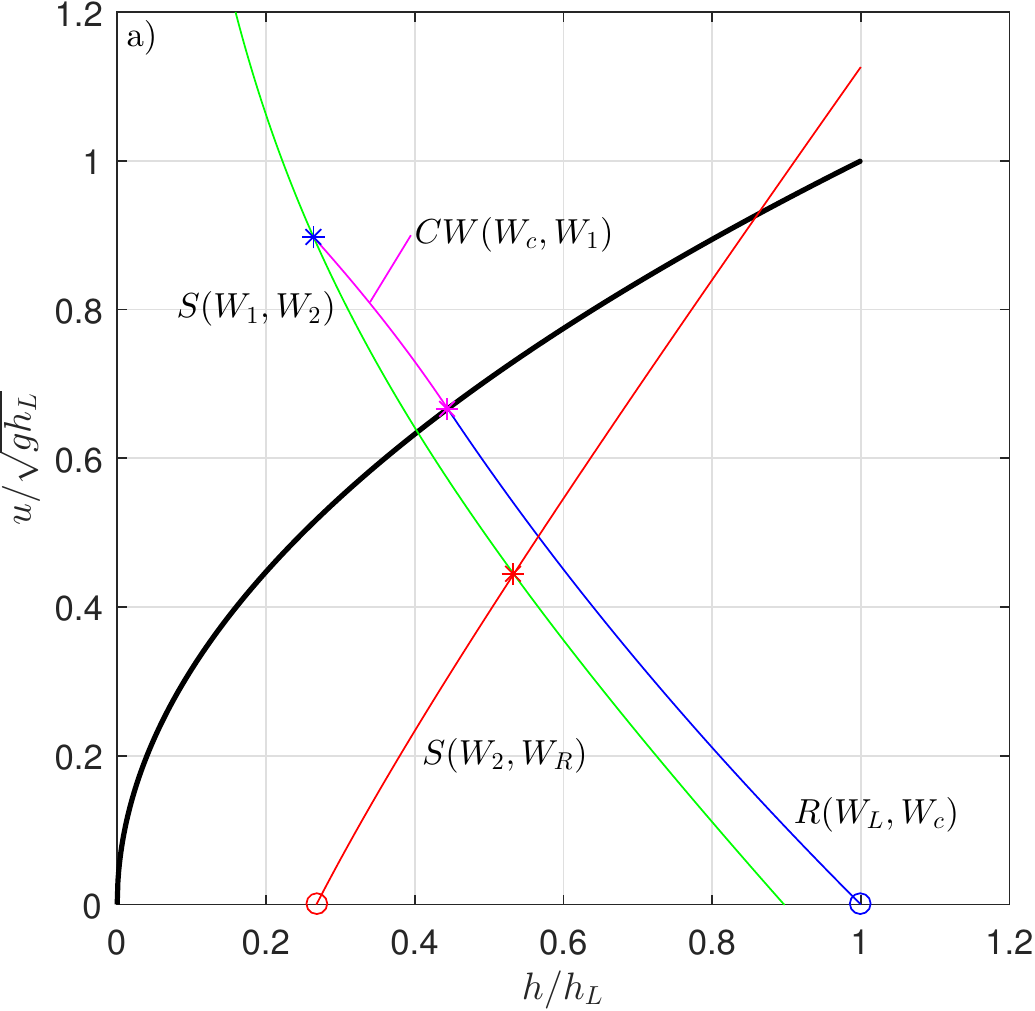}

\rule{0mm}{6mm}

\includegraphics[width=1.0\textwidth]{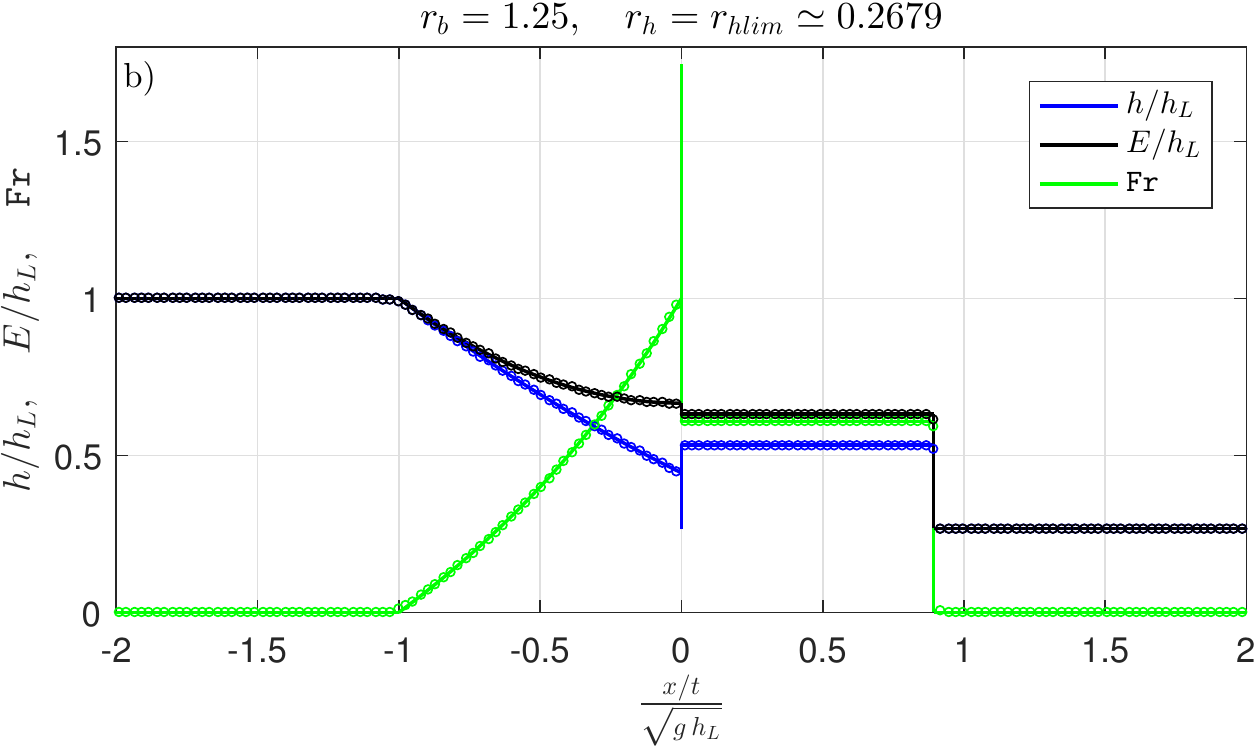}
\end{center}
\caption{Expansion, upper limit depth ratio. a) solution in the phase plane. b) solution in the physical plane. The continuous line is the analytical solution, and the circles represent the numerical solution.}
\label{fig:EULDR}
\end{figure}

Using the classic analytical solution of the Riemann problem (Fig.~\ref{fig:EULDR}a), the reasoning is analogous to that of the previous subsection \ref{subsec:ESDR}. The rarefaction curve $R$ (continuous blue line) is drawn up to the resonance curve. The magenta asterisk identifies the critical point at $0^{-}$. From this point, the contact wave curve $CW$ (continuous magenta line) is drawn up to the point (blue asterisk) where $b=b_R$, where the ($h_1$, $u_1$) state occurs. From this point, the shock curve (continuous green line) is drawn, up to the intersection with the right shock curve (continuous red line) at a point (red asterisk), which identifies the ($h_2$, $u_2$) condition. The blue asterisk and the red asterisk identify the upstream and downstream conjugate depths of the stationary jump, respectively, positioned at the dam.

The reasoning is the same as the previous subsection \ref {subsec:ESDR}, up to Eq.~\eqref{eq:Fr_1_0p}, inclusive. Imposing $c_1 = 0$, from Eq.~\eqref{eq:dc1}, we obtain:
\begin{equation}\label{eq:c1_null}
\Fr_1^2=\frac{u_1^2}{g \, h_1} = \frac{1}{2} \, \left[  \left( \frac{h_2}{h_1} \right) + \left( \frac{h_2}{h_1}\right)^2 \right]
\end{equation}
so that the sequent depth ratio $h_2/h_1$ is the classic ratio for the stationary hydraulic jump:
\begin{equation}\label{eq:h2suh1}
\left(\frac{h_2}{h_1}\right)_{lim} = \frac{1}{2} \, \left( \sqrt{1+8\, \Fr_1^2}-1 \right)  
\end{equation}
The corresponding value of the velocity can be found using Eq.~\eqref{eq:ds1}, once \eqref{eq:h2suh1} gives the value of $h_2$.
It is important to emphasize that in such limit conditions, the length of the constant state ($h_1$, $u_1$) is reduced to zero, and the stationary shock is confined in $x=0^{+}$. Therefore, the physical variables have a double value in $x=0^{+}$.

Eq.s~\eqref{eq:hR2uL} and \eqref{eq:theta} are still valid, giving the necessary relationship between $h_2$ and $h_R$.

In conclusion, the $r_{hlim}$ limit ratio can be found, which divides the \emph{intermediate} depth ratio from the \emph{small} depth ratio in the case of expansion, causing the first shock to collapse in a stationary shock at $x=0^{+}$, as:
\begin{equation}\label{eq:rh_ulim}
r_{hlim}=\left(\frac{h_R}{h_L}\right)_{lim}=\left(\frac{h_R}{h_2}\right) \, \left( \frac{h_2}{h_L} \right)_{lim}=\left(\frac{h_R}{h_2}\right) \, \left( \frac{h_2}{h_1} \right)_{lim} \, \left( \frac{h_1}{h_L} \right)
\end{equation}
where $(h_1/h_L)$ is obtained from Eq.~\eqref{eq:h1suhL}, $(h_2/h_1)_{lim}$ from Eq.~\eqref{eq:h2suh1} and $(h_R/h_2)$ from Eq.~\eqref{eq:hR2uL}.

Referring to Fig.~\ref{fig:CLIMR}, if the point $(r_b, r_h)$ lies over the limit curve and under the first upper limit curve, the solution is an \emph{intermediate} depth ratio solution (magenta area); if the point lies under the same curve and over the lower limit curve, the solution is a \emph{small} depth ratio solution (cyan area).

The typical solution is depicted (continuous line) in the physical plane in Fig.~\ref{fig:EULDR}b, together with the corresponding numerical solution (circles).

\subsection{Expansion, very small depth ratio}
\label{subsec:EXSDR}
A further configuration is possible for a \emph{very small} downstream/upstream dept ratio $r_h$. In such cases, the reasoning is analogous to that in subsection \ref{subsec:ESDR}, up to the constant state described by Eqs.~\eqref{eq:h1suhL}-\eqref{eq:Fr_1_0p}, inclusive. Such a constant state is followed by a rarefaction, up to the constant state ($h_2$, $u_2$). Finally, the downstream shock dividing such a constant state from the final still-water state ($h_R$, $u_R=0$) is of the same kind as that described in subsection \ref{subsec:ESDR}. In practice, the difference between the present case and the \emph{small} $r_h$ case is that the shock moving downstream with a celerity $c_1$ is replaced by a rarefaction.

\begin{figure}
\begin{center}
\includegraphics[width=0.8\textwidth]{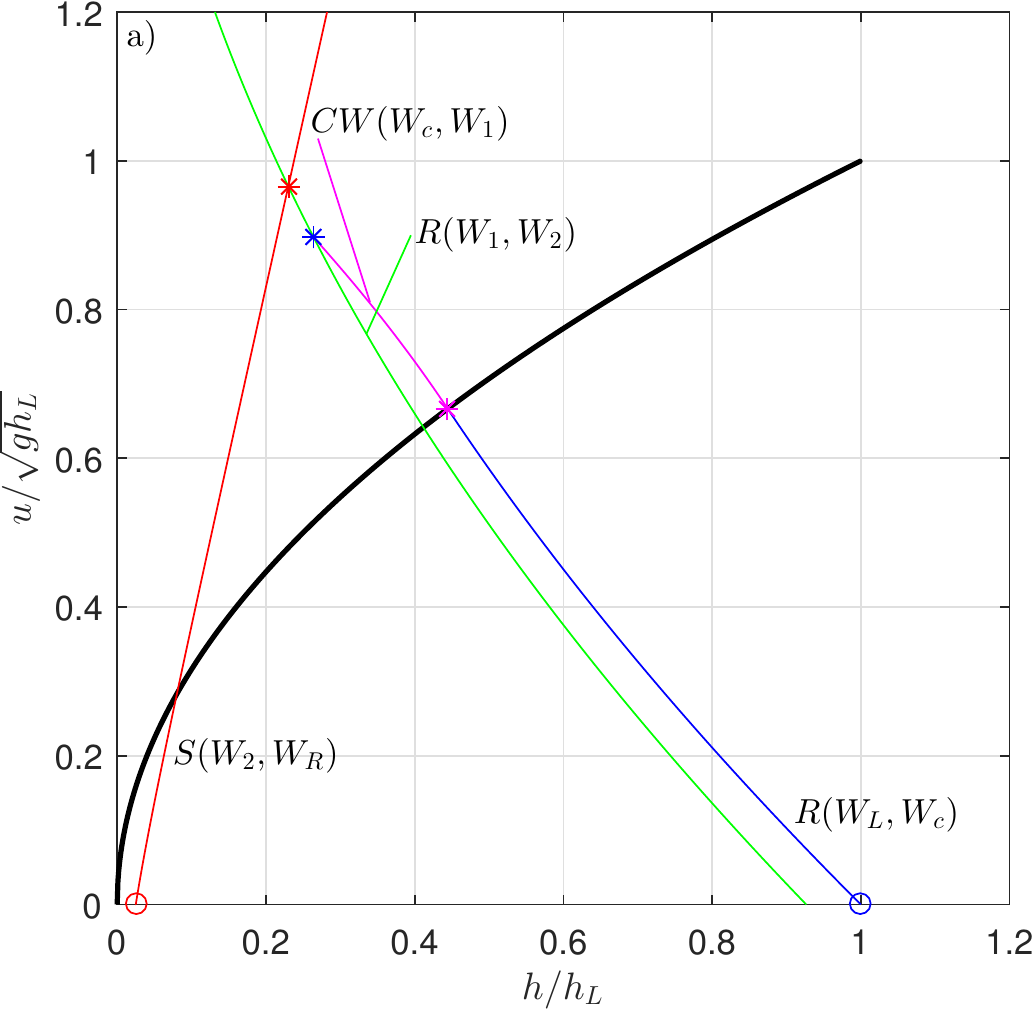}

\rule{0mm}{6mm}

\includegraphics[width=1.0\textwidth]{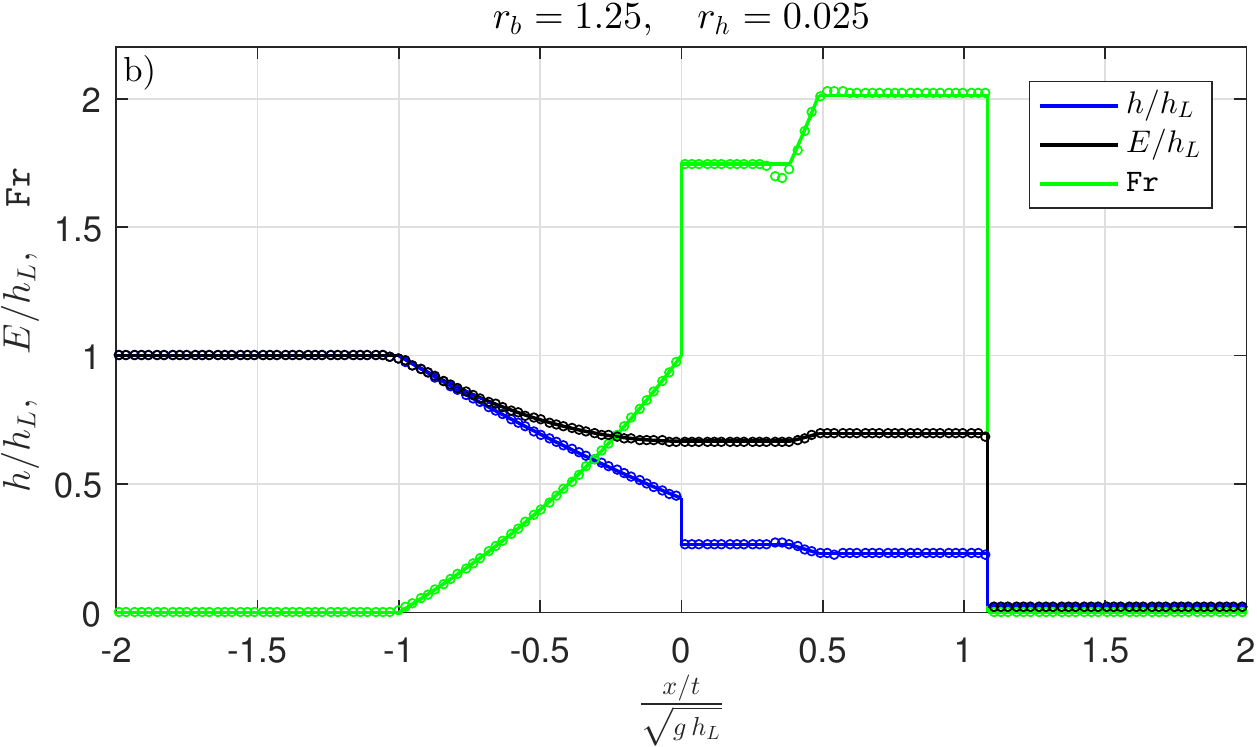}
\end{center}
\caption{Expansion, very small $r_h$. a) solution in the phase plane. b) solution in the physical plane. The continuous line is the analytical solution, and the circles represent the numerical solution.}
\label{fig:EXSDR}
\end{figure}

Using the classic analytical solution of the Riemann problem  (Fig.~\ref{fig:EXSDR}a), the diagram is analogous to that for the \emph{small} $r_h$ case, Fig.~\ref{fig:ESDR}a, and specifically for the rarefaction curve $R$ (continuous blue line), the magenta asterisk identifying the critical point at $0^{-}$, the contact wave curve $CW$ (continuous magenta line) up to the condition $b=b_R$, obtaining the ($h_1$, $u_1$) state (blue asterisk). From this point, a rarefaction curve (continuous green line) is computed instead of the shock curve of Fig.~\ref{fig:ESDR}. Such a curve intersects the shock curve (continuous red line) starting from the right state at a point (red asterisk), which identifies the ($h_2$, $u_2$) condition. The blue asterisk and the red asterisk identify the two constant states, both downstream of the dam, divided by a rarefaction wave.

Moving from $0^{+}$ downstream, the solution is governed by the following system of equations (upstream to the dam position, the nature of the solution is described in the subsection \ref {subsec:ESDR}).

The constant state downstream of the dam position ($x_{r1}$ is the position of the rarefaction head):
\begin{equation}\label{eq:CS1}
\left\{ \begin{array}{l} h = h_1 \\ u = u_1 \end{array} \right . \quad \textrm{for} \quad 0^{+} \leq \frac{x}{t} \leq \frac{x_{r1}}{t} = \left(u_1-\sqrt{g \, h_1}\right)   
\end{equation}

Rarefaction relating two supercritical flows ($x_{r2}$ is the position of the rarefaction tail):
\begin{equation}\label{eq:DREE}
u_1 + 2 \sqrt{g \, h_1} = u + 2 \sqrt{g \, h} \quad \textrm{for} \quad \frac{x_{r1}}{t} \leq \frac{x}{t} \leq \frac{x_{r2}}{t} = \left(u_2-\sqrt{g \, h_2}\right) 
\end{equation}

Constant state downstream of the last rarefaction:
\begin{equation}\label{eq:CS2}
\left\{ \begin{array}{l} h = h_2 \\ u = u_2 \end{array} \right . \quad \textrm{for} \quad \frac{x_{r2}}{t} \leq \frac{x}{t} \leq \frac{x_{2}}{t} = c_2  
\end{equation}
where $c_2$ is again obtained from Eq.~\eqref{eq:c2}.

Final constant state downstream of the shock moving with celerity $c_2$:
\begin{equation}\label{eq:CSREE}
\left\{ \begin{array}{l} h = h_R \\ u = u_R \end{array} \right . \quad \textrm{for} \quad \frac{x}{t} \geq \frac{x_2}{t}   
\end{equation}

The couple $(h_2, u_2)$ is found from the Newton-Raphson solution of the system:
\begin{subequations}\label{eq:sys2}
\begin{align}
&u_1 + 2 \sqrt{g \, h_1} = u_2 + 2 \sqrt{g \, h_2}
\label{eq:sys2a}\\
&u_2 = \left(h_2-h_R\right) \sqrt{\frac{1}{2}\,g\left(\frac{1}{h_2}+\frac{1}{h_R} \right)}
\label{eq:sys2b}
\end{align}
\end{subequations}
since the relation \eqref{eq:RHx2} still holds and $h_1$ and $u_1$ are found from Eqs.~\eqref{eq:h1suhL} and \eqref{eq:Fr_1_0p}, respectively.

The typical solution is depicted (continuous line) in the physical plane in Fig.~\ref{fig:EXSDR}b, together with the corresponding numerical solution (circles).

\subsection{Expansion, lower limit depth ratio}
\label{subsec:ELLDR}
This subsection analytically finds the limit curve dividing a \emph{small} depth ratio from a \emph{very small} depth ratio in the case of expansion to identify the range of existence of both. The analysis investigating such cases, conducted in subsections \ref {subsec:ESDR} and \ref {subsec:EXSDR}, allows us to identify this lower limit depth ratio $r_{hlim}^{*}$, depending on $r_b$, as the only one that provides two identical constant states downstream of the dam, that is $h_1=h_2$; $u_1=u_2$. This result means that there is only one state between $x=0^{+}$ and $x=x_2$. This condition happens for a particular value of $h_R$, once $h_L$ is given. To obtain this result, the reasoning is the same as that of subsection \ref{subsec:ESDR}, evaluating Eq.~\eqref{eq:h1suhL} and Eq.~\eqref{eq:Fr_1_0p}. Referring to subsection \ref{subsec:EULDR}, Eq.~\eqref{eq:c1_null} and Eq.~\eqref{eq:h2suh1} are no longer valid and are replaced by:
\begin{subequations}\label{eq:equality}
\begin{align}
h_2 = h_1\label{eq:equalitya}\\
u_2 = u_1\label{eq:equalityb}
\end{align}
\end{subequations}

\begin{figure}
\begin{center}
\includegraphics[width=0.8\textwidth]{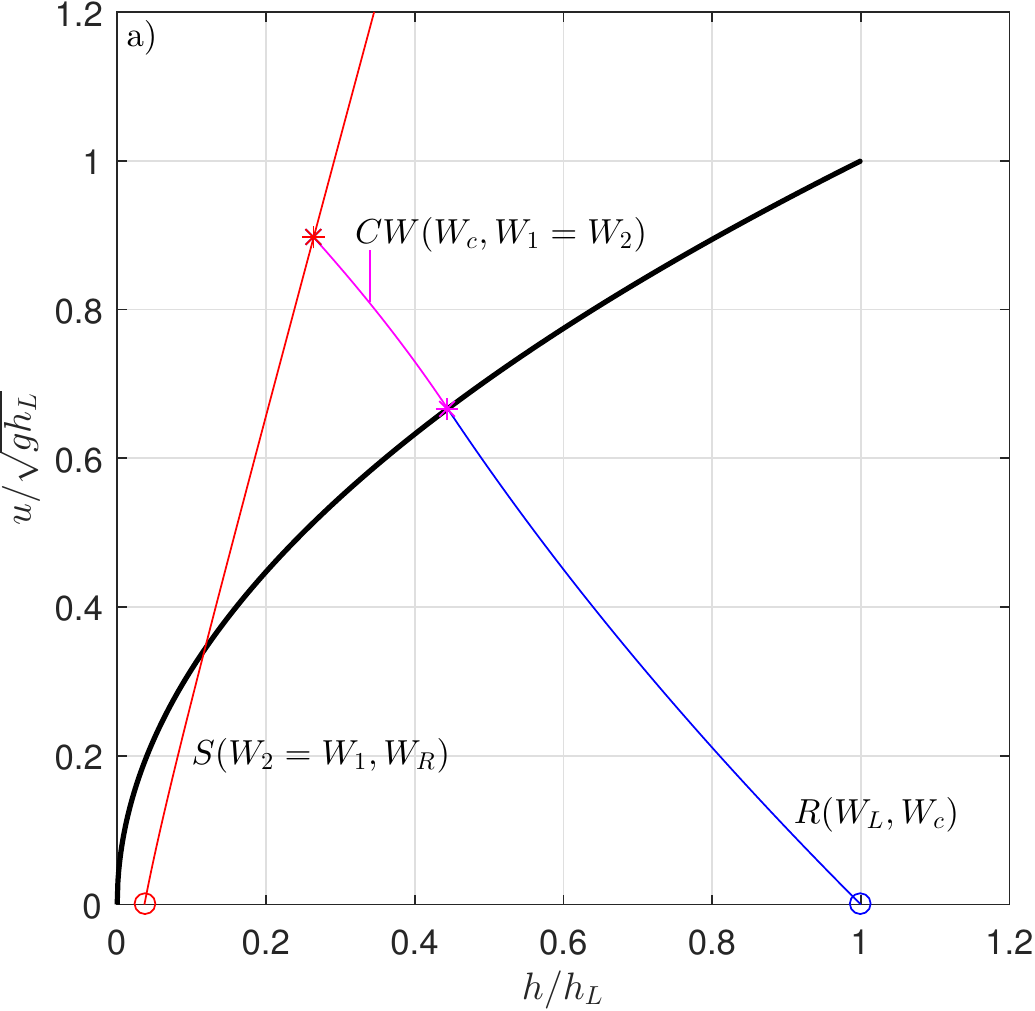}

\rule{0mm}{6mm}

\includegraphics[width=1.0\textwidth]{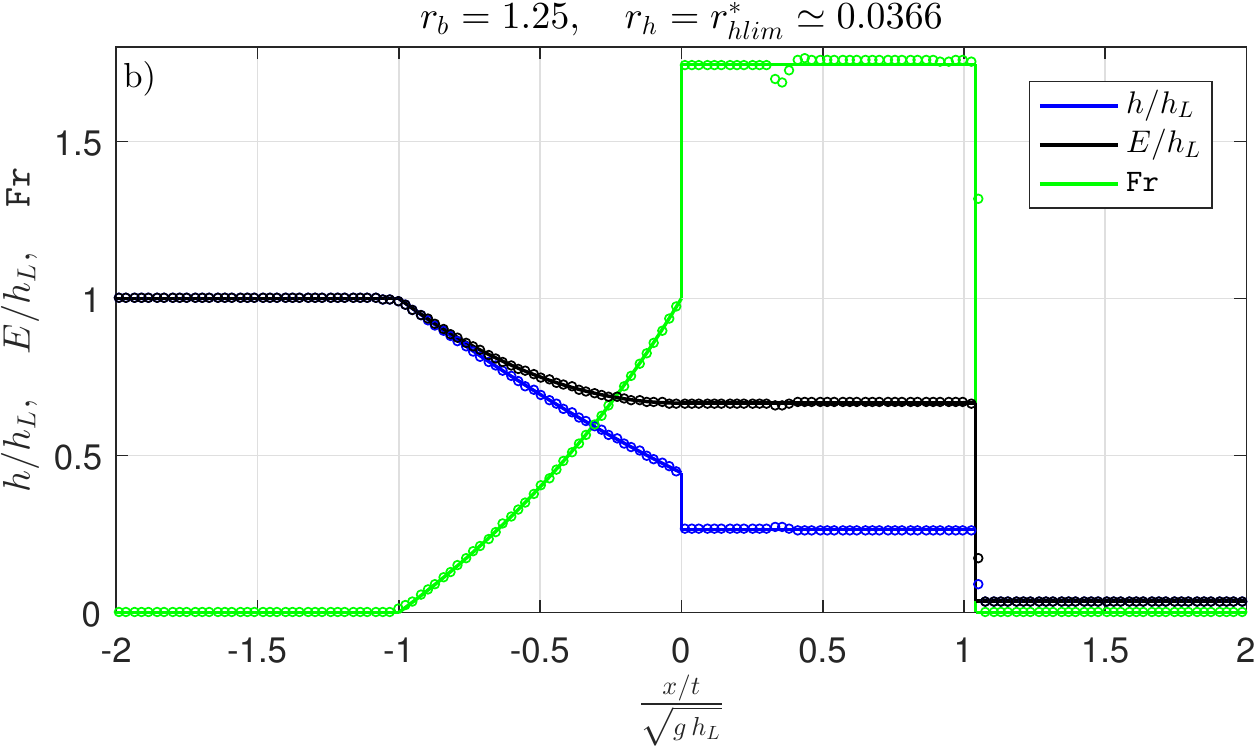}
\end{center}
\caption{Expansion, lower limit depth ratio. a) solution in the phase plane. b) solution in the physical plane. The continuous line is the analytical solution, and the circles represent the numerical solution. }
\label{fig:ELLDR}
\end{figure}

Using the classic analytical solution of the Riemann problem (Fig.~\ref{fig:ELLDR}a), the reasoning is analogous to that of the previous subsection \ref{subsec:EXSDR}. The rarefaction curve $R$ (continuous blue line) is drawn up to the resonance curve. The magenta asterisk identifies the critical point at $0^{-}$. From this point, the contact wave curve $CW$ (continuous magenta line) is computed up to the condition $b=b_R$, obtaining the ($h_1$, $u_1$) state (blue asterisk, which is not visible because it is covered by a red asterisk). The shock curve (continuous red line) from the right state passes through this point, which identifies also the ($h_2$, $u_2$) state (red asterisk exactly on the blue one). The two states 1 and 2 collapse into a unique state, the constant state downstream of the dam.

Once $(h_1, u_1)$ = $(h_2, u_2)$ conditions are found, the procedure is again the same as in subsection \ref{subsec:EULDR}, up to:
\begin{equation}\label{eq:Frcond}
\left(\frac{h_R}{h_1}\right)_{lim}=  \frac{1}{3} - \frac{2}{3} \sqrt{3\left(1+2 \, \Fr_1^2 \right)+1} \, \cos \left( \frac{\theta}{3} + \frac{\pi}{3} \right)
\end{equation}
which is identical to Eq.~\eqref{eq:hR2uL}, with $(h_1, \Fr_1)$ instead of $(h_2, \Fr_2)$.

The conclusion is strictly similar to Eq.~\eqref{eq:rh_ulim}, with the special value $h_2=h_1$ for the depth:
\begin{equation}\label{eq:rh_lim*}
r_{hlim}^{*}=\left(\frac{h_R}{h_L}\right)_{lim}=\left(\frac{h_R}{h_1}\right)_{lim} \, \left( \frac{h_1}{h_L} \right)
\end{equation}

In conclusion, the $r_{hlim}^{*}$ limit ratio can be found, which divides the \emph{small} depth ratio from the \emph{very small} depth ratio in the case of expansion, causing the two constant states downstream of the dam to collapse in a unique state.
Such $r_{hlim}^{*}$ is a function of the $r_b$ value because $h_1/h_L$ depends on $\beta$ (Eq.~\eqref{eq:h1suhL}), which depends in turn on $r_b$, according to Eq.~\eqref{eq:beta_rb}.

\begin{figure}
\begin{center}
\includegraphics[width=1.0\textwidth]{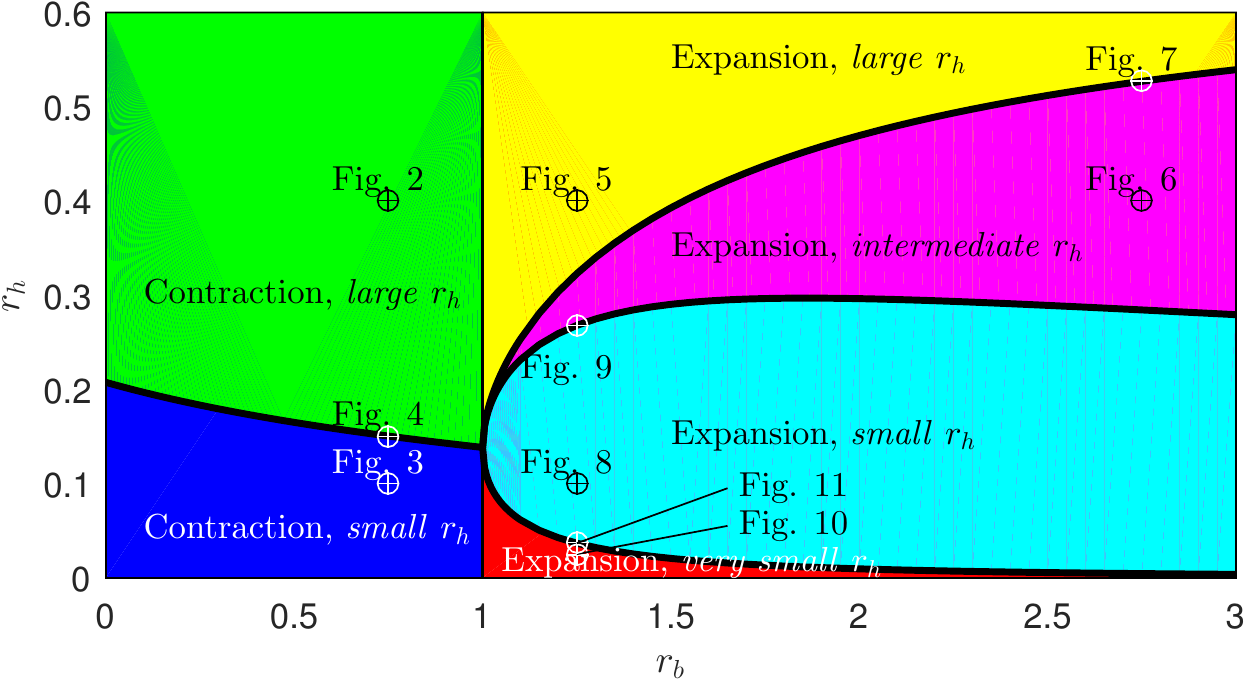}
\end{center}
\caption{Limit curves in the plane ($r_b$, $r_h$). Analytical expressions are derived in subsections: \ref{subsec:CLLDR}, Eq.~\eqref{eq:g_1}; \ref{subsec:EULDR}, Eq.~\eqref{eq:rh_ulim}; \ref{subsec:ELLDR}, Eq.~\eqref{eq:rh_lim*}.}
\label{fig:CLIMR}
\end{figure}

Given a certain width ratio $r_b$ and an initial condition on the depth ratio $r_h$, if the point ($r_b$, $r_h$) lies over this lower limit $r_{hlim}^{*}$ and under the previously defined second upper limit $r_{hlim}$, the configuration of the solution is that of the \emph{small} depth ratio (cyan area in Fig.~\ref{fig:CLIMR}); if such a point ($r_b$, $r_h$) lies under this lower limit $r_{hlim}^{*}$, the configuration is that of the \emph{very small} depth ratio (red area in Fig.~\ref{fig:CLIMR}).

The typical lower limit solution is depicted (continuous line) in the physical plane in Fig.~\ref{fig:ELLDR}b, together with the corresponding numerical solution (circles).

Finally, a complete panorama of the kinds of solutions is now completed and is summarized in Fig.~\ref{fig:CLIMR}.
This diagram can be considered the summary of this work.

\section{Numerical method for channel width discontinuities}
\label{sec:NMCW}
In this section, a suitable numerical method, designed to capture the balancing of the system \eqref{eq:CSWE} also in the case of contact discontinuities, is described.
The physical domain, whose length is $L$, is divided into $N$ cells of size $\Delta x = L/N$.
The $i$-th cell, with $i=1,\ldots, N$, is $I_i= [x_{i+1/2},x_{i-1/2}]$, where $x_{i\pm1/2}$ are the cell boundary positions and $x_{i}$ is its center.
The current time is $t^{n}$, and the time step is $\Delta t = t^{n+1} - t^{n}$. The intermediate time step is defined as $t^{n+1/2} = t^{n} + \Delta t /2$. A second-order FVM in space and time is chosen, as described in \cite{Leibinger, Bertaglia2018}.
The Dumbser-Osher-Toro (DOT) approximate Riemann solver \cite{DOT} is adopted to evaluate fluctuations at the cell boundaries related to the nonconservative part of the system \eqref{eq:CSWE}. The slope of the dependent variables in the $i$-th cell is estimated using the classic minmod slope limiter \cite{Toro99}, as follows:
\begin{equation}\label{eq:minmod}
\Delta W_i^n = \text{minmod} \left(W_i^n-W_{i-1}^n,\, W_{i+1}^n-W_i^n \right)
\end{equation}
where $W_i^n(t)$ is the cell averaged vector of the conservative variables.
A first estimate of the time derivative at time $t^{n}$ is given by:
\begin{equation}\label{eq:dert}
\frac{\d W_i^n}{\d t} = -A(W_i^n)\, \frac{\Delta W_i^n}{\Delta x}
\end{equation}
so that the dependent variable at the cell $i$-th after a one-half time step can be written as:
\begin{equation}\label{eq:dertnpunm}
W_{i}^{n+1/2} = W_i^n + \frac{1}{2}\, \Delta t \,\frac{\d W_i^n}{\d t}
\end{equation}

The application of the DOT method in nonconservative form \cite{Leibinger, Carraro2018} gives the following expression after one time step at time $t^{n+1}$:
\begin{equation}\label{eq:Wagg}
W_i^{n+1} = W_i^n - \frac{\Delta t}{\Delta x}\, \left( D_{i+1/2}^{-}+D_{i-1/2}^{+}\right) - \Delta t \, A(W_i^{n+1/2}) \frac{\Delta W_i^n}{\Delta x}
\end{equation}
Eq.~\eqref{eq:Wagg} requires the evaluation of the fluctuations:
\begin{equation}\label{eq:Fluct}
D_{i+1/2}^{\pm} = \frac{1}{2} \int_{0}^{1}{\left[A(\Psi(s))\pm\left|A(\Psi(s))\right|\right] \frac{\partial \Psi}{\partial s}\, \d s}
\end{equation}
where:
\begin{equation}\label{eq:path}
\Psi(s) = \Psi \left(W_{i+1/2}^{-},\,W_{i+1/2}^{+};\, s \right), \quad 0\le s \le 1
\end{equation}
is a proper path \cite{DLM, Castro2008}, connecting the inner and outer values of the dependent variable at the cell boundaries.
Inside Eq.~\eqref{eq:path}, the dependent variable at the $i$-th cell internal boundaries are estimated as:
\begin{subequations}
\begin{align}
W_{i+1/2}^{-} &= W_i^n + \frac{1}{2} \Delta W_i^n + \frac{1}{2} \Delta t  \frac{\d W_i^n}{\d t}\label{eq:Wmeno}\\
W_{i-1/2}^{+} &= W_i^n - \frac{1}{2} \Delta W_i^n + \frac{1}{2} \Delta t  \frac{\d W_i^n}{\d t}\label{eq:Wpiu}
\end{align}\label{eq:Wpiumeno}
\end{subequations}

The quantity:
\begin{equation}\label{eq:Jac}
|A| = R\, |\Lambda| \, L
\end{equation}
is the absolute value of the $A$ matrix, computed using the absolute values of the three eigenvalues, forming the $\Lambda$ diagonal matrix:
\begin{equation}\label{eq:Leig}
\Lambda = \diag\begin{bmatrix}u-c,&0,&u+c\end{bmatrix}
\end{equation}

The simplest, and more frequently used, path is the linear one \cite{Leibinger, Bertaglia2018}:
\begin{equation}\label{eq:linpath}
\Psi \left(W_{i+1/2}^{-},\,W_{i+1/2}^{+};\, s \right) = W_{i+1/2}^{-} + \left(W_{i+1/2}^{+}-W_{i+1/2}^{-}\right)\,s \, , \quad 0 \le s \le 1
\end{equation}

\begin{figure}
\begin{center}
\includegraphics[width=1.0\textwidth]{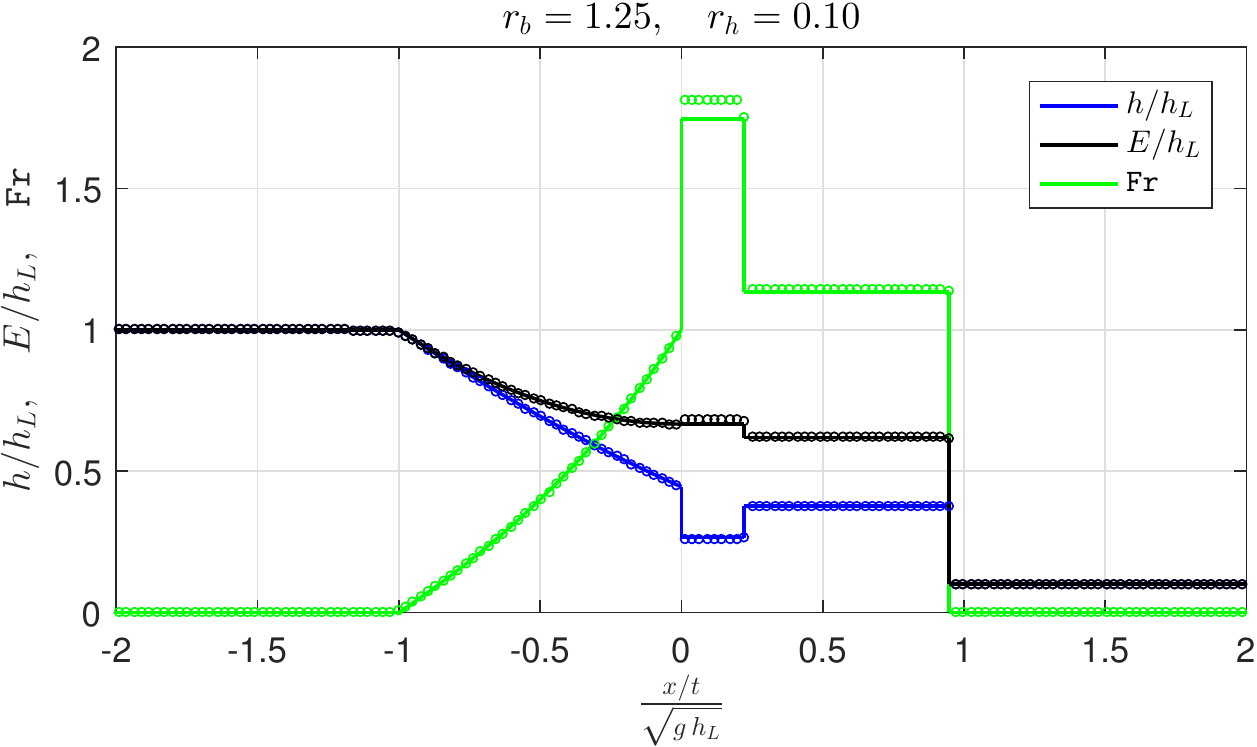}
\end{center}
\caption{Expansion, small depth ratio. The continuous line is the analytical solution, and the circles represent the numerical solution in the case of a linear path in the fundamental variables $[h, q, b]$.}
\label{fig:ESDRLIN}
\end{figure}

Obviously, the linear path also represents the first attempt conducted here to test the model capability in reproducing the analytical results.
As can be deduced from Fig.~\ref{fig:ESDRLIN}, the linear path is not able to capture the energy conservation at the width discontinuity, where a systematic error is introduced. This finding  is not surprising, as the conserved quantities at the singularity are the total discharge and the specific energy, and neither of them is included in the dependent variables. Thus, a different choice is adopted, inspired by the structure of the generalized Riemann invariants, see Eq.~\eqref{eq:GR2}, and previously successfully used in a quite similar context, that is, the shallow water flow over a step \cite{Caleffi2017}. In a different context, a nonlinear path is also used in in blood flow simulation in vessels \cite{Muller2013}.

The proposed path is linear in the variables $Q$, $E$, and $b$ and nonlinear in the variables $q$ and $h$, as follows:
\begin{subequations}\label{eq:nonlinpath}
\begin{align}
&\Psi_Q \left(Q_{i+1/2}^{-},\,Q_{i+1/2}^{+};\, s \right) = Q_{i+1/2}^{-} + \left(Q_{i+1/2}^{+}-Q_{i+1/2}^{-}\right)s
\label{eq:nlQ}\\
&\Psi_E \left(E_{i+1/2}^{-},\,E_{i+1/2}^{+};\, s \right) = E_{i+1/2}^{-} + \left(E_{i+1/2}^{+}-E_{i+1/2}^{-}\right)s
\label{eq:nlE}\\
&\Psi_b \left(b_{i+1/2}^{-},\,b_{i+1/2}^{+};\, s \right) = b_{i+1/2}^{-} + \left(b_{i+1/2}^{+}-b_{i+1/2}^{-}\right)s
\label{eq:blb}
\end{align}
\end{subequations}
for $0 \le s \le 1$.
As a consequence, the nonlinear path for $q$ is:
\begin{equation}\label{eq:pathq}
q\left(s\right) = Q\left(s\right)/b\left(s\right)  \, , \quad 0 \le s \le 1
\end{equation}
and the nonlinear path for $h(s)$ is implicitly defined from:
\begin{equation}\label{eq:pathh}
h\left(s\right) + \frac{\left[q\left(s\right)\right]^2}{2 \, g \, \left[h\left(s\right)\right]^2} =E\left(s\right)  \, , \quad 0 \le s \le 1
\end{equation}
Equation \eqref{eq:pathq} can be used immediately, whereas equation \eqref{eq:pathh} requires the inversion of a 3rd-degree equation, which usually admits two real positive solutions, one subcritical (corresponding to $\Fr <1$) and one supercritical (corresponding to $\Fr >1$), which must be selected on the basis of the flow Froude number at the cell boundaries ($s = 0$ and $s = 1$). The analytical inversion of Eq.~\eqref{eq:pathh} can be found in \cite{Valiani2008} and \cite{Valiani2017} and is extensively used in different numerical applications \cite{Caleffi2016, Caleffi2017}. The condition to be satisfied to obtain that Eq.~\eqref{eq:pathh} can be inverted is:
\begin{equation}\label{eq:energyineq}
E\left(s\right) \ge E_c\left(s\right) = \sqrt[3]{\frac{\left[q\left(s\right)\right]^2}{g}}
\end{equation}
When condition \eqref{eq:energyineq} is not satisfied or when the quantity $\Fr^2-1$ has different signs across the cell boundary, the simple linear path \eqref{eq:linpath} is locally adopted. A more refined, but more complex, nonlinear structure of the path is investigated for geometrical singularities consisting of bed steps in \cite{Caleffi2016, Caleffi2017}.

Concerning the numerical quadrature of the integrals in Eq.~\eqref{eq:Fluct}, a Gauss-Legendre formula is used.
A three-point quadrature is verified to be consistent with the chosen order of accuracy of the numerical method.

The codes, written in MATLAB (MathWorks Inc.) language, are made available in the Mendeley Data repository associated to this article \cite{This}.

\section{Conclusions}
\label{sec:Concl}
The complete solution of the dam break over a wet bed in channels where upstream and downstream cross-section widths are different is given. Two solution configurations are found for channel contractions, and four solution configurations are found for channel expansions. The phenomenon is governed by the two nondimensional parameters $r_b$ and $r_h$. In the plane ($r_b$, $r_h$), the limit curve dividing the \emph{large} depth ratio and the \emph{small} depth ratio in the case of contraction is analytically found. Moreover, the three limit curves dividing the \emph{large} depth ratio, the \emph{intermediate} depth ratio, the \emph{small} depth ratio and the \emph{very small} depth ratio in the case of expansion are analytically found.

The plane ($r_b$, $r_h$) is consequently divided into six regions, each of which is associated with a precise configuration of the solution. Four of these six regions are associated with resonant cases, giving an overall behavior that is considerably richer than the classic constant-width Stoker dam-break problem.

A second-order Dumbser-Osher-Toro numerical method, equipped with a nonlinear path connecting discontinuous values at cell boundaries, is shown to be able to capture the main flow features in all possible configurations. In particular, incorporating a nonlinear path inspired by the structure of generalized Riemann invariants allows proper mass conservation and specific energy conservation across the contact wave occurring at the dam position.

This method appears to be promising for treating nonconservative balance law formulations when geometrical singularities occur.

The collection of analytical results can also be used to consolidate an analytical database for the proper validation of numerical methods to apply to shallow water or similar balance laws when significant contact waves require a focused treatment.

\section{Acknowledgments}
\label{sec:Ack}
This work was supported by FAR 2018 (University of Ferrara) and by PRIN 2017 grant, code 2017KKJP4X MIUR (Ministero dell'Istruzione, della Universit\`a e della Ricerca).


\begin{thebibliography}{10}
\expandafter\ifx\csname url\endcsname\relax
  \def\url#1{\texttt{#1}}\fi
\expandafter\ifx\csname urlprefix\endcsname\relax\def\urlprefix{URL }\fi
\expandafter\ifx\csname href\endcsname\relax
  \def\href#1#2{#2} \def\path#1{#1}\fi

\bibitem{Stoker}
J.~J. Stoker, Water waves, Interscience Publishers, Wiley and Sons, New York,
  1957.

\bibitem{Henderson1966}
F.~M. Henderson, Open Channel Flow, Macmillan, 1966.

\bibitem{Liggett}
J.~A. Liggett, {Fluid mechanics}, McGraw-Hill Inc., New York, 1994.

\bibitem{Chaudry}
M.~H. Chaudry, Open-Channel Flow, second edition Edition, Springer, 2008.

\bibitem{Cadam}
M.~Morris, J.~C. Galland, P.~Balabanis, {Concerted Action on Dam-break
  modelling}, in: E.~Union (Ed.), Proceedings of the CADAM meeting Wallingford,
  UK, European Union, Office for Official Publications of the European
  Communities, 1999, Rue de la Loi/Wetstraat 200, B-1049 Brussels, 1998.

\bibitem{Valiani2002}
A.~Valiani, V.~Caleffi, A.~Zanni, {Case Study: Malpasset Dam-Break Simulation
  Using a Two-Dimensional Finite Volume Method}, ASCE Journal of Hydraulic
  Engineering 128~(5) (2002) 460--472.

\bibitem{Caleffi2003}
V.~Caleffi, A.~Valiani, A.~Zanni, {Finite Volume Method for Simulating Extreme
  Flood Events in Natural Channels}, IAHR Journal of Hydraulic Research 41~(2)
  (2003) 167--177.

\bibitem{Singh2011}
J.~Singh, M.~S. Altinakar, Y.~Ding, Two-dimensional numerical modeling of
  dam-break flows over natural terrain using a central explicit scheme,
  Advances in Water Resources 34~(10) (2011) 1366--1375.

\bibitem{Wang2011}
Y.~Wang, Q.~Liang, G.~Kesserwani, J.~W. Hall, A 2{D} shallow flow model for
  practical dam-break simulations, Journal of Hydraulic Research 49~(3) (2011)
  307--316.

\bibitem{alcrudo01}
F.~Alcrudo, F.~Benkhaldoun, {Exact solutions to the Riemann problem of the
  shallow water equations with a bottom step}, Computers and Fluids 30~(6)
  (2001) 643--671.

\bibitem{Valiani2017}
A.~Valiani, V.~Caleffi, Momentum balance in the shallow water equations on
  bottom discontinuities, Advances in Water Resources 100 (2017) 1--13.

\bibitem{LeFloch2011}
P.~G. LeFloch, M.~D. Thanh, A {G}odunov-type method for the shallow water
  equations with discontinuous topography in the resonant regime, Journal of
  Computational Physics 230~(20) (2011) 7631--7660.

\bibitem{LeFloch2007}
P.~G. LeFloch, M.~D. Thanh, The {R}iemann problem for the shallow water
  equations with discontinuous topography, Communications in Mathematical
  Sciences 5~(4) (2007) 865--885.

\bibitem{HW2014}
E.~Han, G.~Warnecke, Exact riemann solutions to shallow water equations,
  Quarterly of Applied Mathematics 72 (2014) 407--453.

\bibitem{Murillo2013}
J.~Murillo, P.~Garc\'ia-Navarro, Energy balance numerical schemes for shallow
  water equations with discontinuous topography, {Journal of Computational
  Physics} 236 (2013) 119--142.

\bibitem{Murillo2014}
J.~Murillo, P.~Garc\'ia-Navarro, Accurate numerical modeling of 1{D} flow in
  channels with arbitrary shape. {A}pplication of the energy balanced property,
  Journal of Computational Physics 260 (2014) 222--248.

\bibitem{Navas2015}
A.~Navas-Montilla, J.~Murillo, {Energy balanced numerical schemes with very
  high order. The Augmented Roe Flux ADER scheme. Application to the shallow
  water equations}, Journal of Computational Physics 290 (2015) 188--218.

\bibitem{Caleffi2016}
V.~Caleffi, G.~Li, A.~Valiani, {A comparison between bottom-discontinuity
  numerical treatments in the DG framework}, Applied Mathematical Modelling
  40~(17--18) (2016) 7516--7531.

\bibitem{Caleffi2017}
V.~Caleffi, A.~Valiani, Well balancing of the {SWE} schemes for moving-water
  steady flows, Journal of Computational Physics 342 (2017) 85--116.

\bibitem{BeVa-94}
A.~Bermudez, M.~E. V\'{a}zquez-Cend\'{o}n, {Upwind Methods for Hyperbolic
  Conservation Laws with Source Terms}, Computers and Fluids 23~(8) (1994)
  1049--1071.

\bibitem{Caleffi2009}
V.~Caleffi, A.~Valiani, {Well-balanced bottom discontinuities treatment for
  high-order shallow water equations WENO scheme}, ASCE Journal of Engineering
  Mechanics 135~(7) (2009) 684--696.

\bibitem{Cozzolino2018}
L.~Cozzolino, V.~Pepe, L.~Cimorelli, A.~D'Aniello, R.~Della~Morte, D.~Pianese,
  The solution of the dam-break problem in the porous shallow water equations,
  Advances in Water Resources 114 (2018) 83--101.

\bibitem{PCL}
C.~Par\'{e}s, Numerical methods for nonconservative hyperbolic systems: A
  theoretical framework, SIAM Journal on Numerical Analysis 44~(1) (2006)
  300--321.

\bibitem{DLM}
G.~Dal~Maso, P.~LeFloch, F.~Murat, Definition and weak stability of
  nonconservative products, Journal de Math\'{e}matiques Pures et
  Appliqu\'{e}es 74 (1995) 483--548.

\bibitem{DOT}
M.~Dumbser, E.~F. Toro, {A Simple Extension of the Osher Riemann Solver to
  Non-conservative Hyperbolic Systems}, Journal of Scientific Computing
  48~(1-3) (2011) 70--88.

\bibitem{Cozzolino2017}
L.~Cozzolino, V.~Pepe, F.~Morlando, L.~Cimorelli, A.~D'Aniello, R.~Della~Morte,
  D.~Pianese, Exact solution of the dam-break problem for constrictions and
  obstructions in constant width rectangular channels, Journal of Hydraulic
  Engineering 143~(11) (2017) 04017047--1--15.

\bibitem{Leibinger}
J.~Leibinger, M.~Dumbser, U.~Iben, I.~Wayand, {A path-conservative Osher-type
  scheme for axially symmetric compressible flows in flexible viscoelastic
  tubes}, Applied Numerical Mathematics 105 (2016) 47--63.

\bibitem{Bertaglia2018}
G.~Bertaglia, M.~Ioriatti, A.~Valiani, M.~Dumbser, V.~Caleffi, Numerical
  methods for hydraulic transients in visco-elastic pipes, Journal of Fluids
  and Structures 81 (2018) 230--254.

\bibitem{Toro99}
E.~F. Toro, {Riemann Solvers and Numerical Methods for Fluid Dynamics},
  Springer-Verlag, 1999.

\bibitem{Carraro2018}
F.~Carraro, A.~Valiani, V.~Caleffi, {Efficient analytical implementation of the
  DOT Riemann solver for the de Saint Venant-Exner morphodynamic model},
  Advances in Water Resources 113 (2018) 189--201.

\bibitem{Castro2008}
M.~J. Castro, P.~G. LeFloch, M.~L. Mu{\~{n}}oz-Ruiz, C.~Par\'{e}s, Why many
  theories of shock waves are necessary: Convergence error in formally
  path-consistent schemes, Journal of Computational Physics 227~(17) (2008)
  8107--8129.

\bibitem{Muller2013}
L.~O. M\"{u}ller, E.~F. Toro, Well-balanced high-order solver for blood flow in
  networks of vessels with variable properties, International Journal for
  Numerical Methods in Biomedical Engineering 29~(12) (2013) 1388--1411.

\bibitem{Valiani2008}
A.~Valiani, V.~Caleffi, {Depth-energy and depth-force relationships in open
  channel flows: Analytical findings}, Advances in Water Resources 31~(3)
  (2008) 447--454.

\bibitem{This}
A.~Valiani, V.~Caleffi, {Data for: Dam break in rectangular channels with different 
upstream-downstream widths}, Mendeley Data (2019), V2, doi: 10.17632/33zc86d8bf.2.

\end{thebibliography}

\end{document}